\documentclass[journal,onecolumn,draftclsnofoot]{IEEEtran}
\usepackage{setspace,verbatim,amsfonts,graphicx,amsmath,amsbsy,amssymb,citesort,epsfig,epsf}
\newtheorem{theorem}{Theorem}[section]
\newtheorem{corollary}[theorem]{Corollary}
\newtheorem{lemma}[theorem]{Lemma}

\newtheorem{definition}{Definition}[section]

{}

\renewcommand{\theequation}{\thesection.\arabic{equation}}

\newfont{\smallmathfont}{cmmib10 at 8pt}

\newcommand{\Db}{{\mathbf D}}

\newcommand{\ab}{{\mathbf a}}
\newcommand{\bb}{{\mathbf b}}
\newcommand{\cb}{{\mathbf c}}
\newcommand{\db}{{\mathbf d}}
\newcommand{\eb}{{\mathbf e}}

\newcommand{\gb}{{\mathbf g}}

\newcommand{\nb}{{\mathbf n}}

\newcommand{\qb}{{\mathbf q}}

\newcommand{\ub}{{\mathbf u}}
\newcommand{\vb}{{\mathbf v}}
\newcommand{\wb}{{\mathbf w}}
\newcommand{\xb}{{\mathbf x}}
\newcommand{\yb}{{\mathbf y}}
\newcommand{\zb}{{\mathbf z}}

\newcommand{\Rd}{{\mathbb R}}

\begin{document}

\title{Compressive MUSIC: A Missing Link between Compressive Sensing and Array Signal Processing}
\author{Jong Min Kim, Ok Kyun Lee, and Jong Chul Ye
\thanks{Parts of this work were presented on 14/04/2010 at the SIAM Conference on
Imaging Science, Chicago, USA, with the title ``Multiple measurement
vector problem with subspace-based algorithm''. The authors are with the Dept. of Bio and Brain Engineering,
Korea Adv. Inst. of Science \& Technology (KAIST), Republic of
Korea. Send correspondence to jong.ye@kaist.ac.kr.}}

 \maketitle

\begin{abstract}
\setstretch{1}
The multiple measurement vector (MMV) problem addresses the identification
of unknown input vectors that share common sparse support. Even
though MMV problems have been traditionally addressed within the
context of sensor array signal processing, the recent  trend is
to apply compressive sensing (CS)  due to its capability to
estimate sparse support even with an insufficient number of snapshots,
in which case classical array signal processing  fails.
However, CS  guarantees the accurate recovery
in a probabilistic manner, which often shows inferior performance in
the regime where the traditional array signal processing approaches
succeed. The apparent dichotomy between the {\em
probabilistic} CS and {\em deterministic} sensor
array signal processing  has not been fully understood. The main
contribution of the present article is a unified
approach that unveils a {missing link} between CS
and array signal processing. The new algorithm, which we call {\em compressive
MUSIC}, identifies the parts of support using CS,
after which the remaining supports are estimated using a novel
generalized MUSIC criterion. Using a large system MMV model, we
show that our compressive MUSIC requires a smaller
number of sensor elements for accurate support recovery than
the existing CS methods and that it can approach the optimal
$l_0$-bound  with finite number of snapshots.
\end{abstract}


\noindent Correspondence to:\\
Jong Chul Ye, Ph.D\\
Associate Professor\\
Department of Bio and Brain Engineering\\
Korea Adv. Inst. of Science and Technology (KAIST)\\
373-1 Guseong-Dong, Yuseong-Gu, Daejon 305-701, Korea\\
Tel: +82-42-350-4320\\
Email: jong.ye@kaist.ac.kr


\noindent {\bf Submitted to IEEE Trans. on Information Theory}


\section{Introduction}

{C}{ompressive} sensing (CS) theory \cite{Do06,CaRoTa06,CaTa05}
addresses the accurate recovery of unknown sparse signals from
underdetermined linear measurements and has become one of the main
research topics in the signal processing area. Compressive sensing has had a
significant impact on many applications, such as magnetic resonance
imaging \cite{jung2007improved,ye2007projection,jung2009kt}, x-ray
computed tomography \cite{chen2008prior}, communication
\cite{Cotter2002scemp}, remote sensing \cite{wagadarikar2008single},
etc. Most of the compressive sensing theories have been developed
to address the single measurement vector (SMV) problem
\cite{Do06,CaRoTa06,CaTa05}. More specifically, let $m$ and $n$  be
positive integers such that $m<n$. Then, the SMV compressive sensing
problem is given by
\begin{eqnarray}\label{eq:y=Ax0}
(P0): & {\rm minimize}~~~\|\xb\|_0\\
~~~& {\rm subject~to}~~~\bb=A\xb, \notag
\end{eqnarray}
where $\bb \in \mathbb{R}^{m}$, $A \in \Rd^{m \times n}$, $\xb \in
\Rd^{n}$, and $\|\xb\|_0$ denotes the number of non-zero elements in
the vector $\xb$. Since (P0) requires  a computationally expensive
combinatorial optimization,   greedy methods \cite{Tr06}, reweighted
norm algorithms \cite{gorodnitsky1997sparse,candes2008enhancing},
convex relaxation using $l_1$ norm \cite{ChDoSa99,CaRoTa06}, or
Bayesian approaches \cite{ji2008bayesian,wipf2004sparse} have been
widely investigated as alternatives. One of the important
theoretical tools within this context is the so-called
restricted isometry property (RIP), which enables us to guarantee the robust recovery of certain input
signals \cite{CaTa05}. More specifically,
a sensing matrix $A\in\mathbb{R}^{m\times n}$ is said to have a
$k$-restricted isometry property(RIP) if there is a constant
$0\leq \delta_k<1$ such that
$$(1-\delta_k)\|\mathbf{x}\|^2\leq \|A\mathbf{x}\|^2\leq (1+\delta_k)\|\mathbf{x}\|^2$$
for all $\mathbf{x}\in\mathbb{R}^n$ such that $\|\mathbf{x}\|_0\leq
k$. It has been demonstrated that $\delta_{2k}<\sqrt{2}-1$ is
sufficient for $l_1/l_0$ equivalence \cite{CaRoTa06}. For many classes
of random matrices, the RIP condition is satisfied with extremely
high probability if the number of measurements satisfies
 $m\geq c k\log (n/k)$ for some constant $c>0$ \cite{CaTa05}.
Ever since the pioneering work by Cand\`{e}s, Romberg, and Tao
\cite{CaRoTa06} was published, many important theoretical discoveries
have been made. For example, the necessary and/or sufficient
conditions for the sparse recovery by maximum likelihood method
\cite{fletcher2009necessary}, $p$-thresholding
\cite{fletcher2009necessary}, and orthogonal matching pursuit
\cite{fletcher-orthogonal} have been extensively studied. Furthermore,
the geometry of $l_1$ recovery has been revealed using the high
dimensional polytope geometry \cite{Donoho2005pnas}. A recent
breakthrough in SMV compressive sensing is the discovery of
an approximate message passing algorithm \cite{donoho2009message} that
has striking similarity with the iterative thresholding method
\cite{daubechies2003iterative}, while achieving theoretical
optimality.

Another important area of compressive sensing research is the
so-called multiple measurement vector problem (MMV)
\cite{chen2006trs,cotter2005ssl,Mishali08rembo,Berg09jrmm}. The MMV
problem addresses the recovery of a set of sparse signal vectors
that share common non-zero support. More specifically, let $m$, $n$
and $r$ be positive integers such that $m<n$. In the MMV context,
$m$ and $r$ denote the number of sensor elements and snapshots,
respectively.  For a given observation matrix
$B\in\mathbb{R}^{m\times r}$, a sensing matrix $A\in
\mathbb{R}^{m\times n}$ such that $B=AX_*$ for some $X_*\in
\mathbb{R}^{n\times r}$, the multiple measurement vector (MMV)
problem is formulated as:
\begin{eqnarray}\label{eqdefmmv}
{\rm minimize}~~~\|X\|_0\\
{\rm subject~to}~~~B=AX, \notag
\end{eqnarray}
where $X=[\mathbf{x}_1,\cdots,\mathbf{x}_r]\in\mathbb{R}^{n\times
r}$ and $\|X\|_0=|{\rm supp}X|$, where ${\rm supp}X=\{1\leq i\leq n
: \mathbf{x}^i\neq 0\}$ and $\mathbf{x}^i$ is the $i$-th row of $X$.
The MMV problem also has many important applications such as
distributed compressive sensing \cite{baron2005distributed},
direction-of-arrival estimation in radar \cite{krim1996two},
magnetic resonance imaging with multiple coils
\cite{pruessmann1999sense}, diffuse optical tomography using
multiple illumination patterns
\cite{joshi2006fully,LeeKimBreYe2010}, etc. Currently, greedy
algorithms such as S-OMP (simultaneous orthogonal matching pursuit)
\cite{tropp2006ass,chen2006trs},  convex relaxation methods using
mixed norm \cite{malioutov2005ssr,tropp2006algorithms}, M-FOCUSS
\cite{cotter2005ssl}, M-SBL (Multiple Sparse Bayesian Learning)
\cite{wipf2006bayesian}, randomized algorithms such as REduce MMV
and BOost (ReMBo)\cite{Mishali08rembo}, and model-based compressive
sensing using block-sparsity
\cite{eldar2009compressed,baraniuk2010model} have also been applied
to the MMV problem within the context of compressive sensing.

In MMV, thanks to the common sparse support, it is quite predictable
that the recoverable sparsity level may increase with the
increasing number of measurement vectors. More specifically, given a
sensing matrix $A$, let ${\rm spark}(A)$ denote the smallest number
of linearly dependent columns of $A$. Then, according to Chen and
Huo \cite{chen2006trs}, Feng and Bresler \cite{Feng97}, if $X\in\mathbb{R}^{n\times r}$ satisfies
$AX=B$ and
\begin{equation}\label{l0-bound-mmv}
\|X\|_0<  \frac{{\rm spark}(A)+{\rm rank}(B)-1}{2} \leq {\rm
spark}(A)-1,
\end{equation}
then $X$ is the unique solution of (\ref{eqdefmmv}). In
\eqref{l0-bound-mmv}, the last inequality comes from the observation
that ${\rm rank}(B)\leq \|X_*\|_0:=|{\rm supp}X_*|$. Recently,
Davies and Eldar showed that (\ref{l0-bound-mmv}) is indeed a
necessary codition for $X$ to be a unique solution for $AX=B$
\cite{DaviesEldar2010}.
 Compared to
the SMV case (${\rm rank}(B)=1$), \eqref{l0-bound-mmv} informs us
that the recoverable sparsity level increases with the number of measurement vectors. Furthermore, average case analysis \cite{eldar2010average} and information theoretic analysis \cite{tang2009performance} have indicated the performance
improvements of MMV algorithms with an increasing number of snapshots.
However,  the performance of the aforementioned MMV compressive sensing algorithms  are not generally satisfactory, and
significant performance gaps still exist from \eqref{l0-bound-mmv}  even for
a noiseless case when only a finite number of snapshots is available.

On the other hand, before the advance of compressive sensing,  the
MMV problem \eqref{eqdefmmv}, which was often termed as
direction-of-arrival (DOA) or the bearing estimation problem, had
been addressed using sensor array signal processing techniques
\cite{krim1996two}. One of the most popular and successful DOA
estimation algorithms is the so-called the MUSIC (MUltiple SIgnal
Classification) algorithm \cite{schmidt1986multiple}. MUSIC first
calculates the signal subspace and noise subspace by decomposing the
empirical covariance matrix; then, by exploiting the orthogonality
between the noise subspace and signal manifold at the correct target
locations, MUSIC identifies the target locations. The MUSIC
estimator has been proven to be a large snapshot (for $r\gg 1$)
realization of the maximum likelihood estimator for any $m>k$, if
and only if the signals are uncorrelated \cite{StNe89}. As will be
shown later  when ${\rm rank}(B)= k$ and the row vectors $X$ are in
general position, the maximum sparsity level that is uniquely
recoverable using the MUSIC approach is
\begin{equation}\label{l0-bound-music}
\|X\|_0< {\rm spark}(A)-1 \ ,
\end{equation}
which implies that the MUSIC algorithm achieves the $l_0$ bound
\eqref{l0-bound-mmv}  of  the MMV when ${\rm rank}(B) = k $. However, one
of the main limitations of the MUSIC algorithm is its failure when ${\rm
rank}(B)< k$. This problem is often called the ``coherent source''
problem within the sensor array signal processing context
\cite{krim1996two}. For example, MUSIC cannot identify any target
with a single snapshot, whereas the compressive sensing approaches
can identify the location with extremely large probability.

To the best of our knowledge, this apparent ``missing link'' between
compressive sensing  and sensor array signal processing for the MMV
problem has not yet been discussed. The main contribution of the
present article is, therefore, to provide a new class of algorithms
that unveils the missing link.  The new algorithm, termed {\em
compressive MUSIC} (CS-MUSIC), can be regarded  as a deterministic
extension of compressive sensing to achieve the $l_0$ optimality, or
as a generalization of the MUSIC algorithm using a probabilistic
setup to address the difficult problem of the coherent sources
estimation. This generalization is due to our novel discovery of a
{\em  generalized MUSIC criterion}, which tells us that an unknown
support of size ${\rm rank}(B)$  can be estimated {\em
deterministically} as long as a $k-{\rm rank}(B)$ support can be
estimated with any compressive sensing algorithm such as S-OMP or
thresholding. Therefore, as ${\rm rank}(B)$ approaches $k$, our
compressive MUSIC approaches the classical MUSIC estimator; whereas,
as ${\rm rank}(B)$ becomes $1$, the algorithm approaches to a
classical SMV compressive sensing algorithm. Furthermore, even if
the sparsity level is not known {\em a priori}, compressive MUSIC
can accurately estimate the sparsity level using the generalized
MUSIC criterion. This emphasizes the practical usefulness of the new
algorithm.
 Since the fraction of the support that should be estimated
probabilistically is reduced from $k$ to $k-{\rm rank}(B)$, one can
conject that the required number of sensor elements for
compressive MUSIC is significantly smaller than that for
conventional compressive sensing. Using the large system MMV model,
we derive  explicit expressions for the minimum number of sensor
elements, which confirms our conjecture. Furthermore, we derive an
explicit expression of the minimum SNR to guarantee the success of
compressive MUSIC. Numerical experiments confirm out theoretical findings.


 The remainder of the paper is organized as
follows. We provide the problem formulation and mathematical
preliminaries in Section~\ref{sec:formulation}, followed by a review
of existing MMV algorithms  in Section~\ref{sec:review}.
Section~\ref{sec:gmusic} gives a detailed presentation of  the
generalized MUSIC criterion, and the required number of sensor
elements in CS-MUSIC
is calculated in Section~\ref{sec:no}. Numerical solutions are given
in Section~\ref{sec:simulation}, followed by the discussion and
conclusion in Section \ref{sec:dis} and ~\ref{sec:conclusion},
respectively.

\section{Problem Formulation and Mathematical Preliminaries}
\label{sec:formulation}

Throughout the paper, $\xb^i$ and $\xb_j$ correspond to the $i$-th
row and the $j$-th column of matrix $X$, respectively. When $S$ is an
index set, $X^S$, $A_S$ corresponds to a submatrix collecting
corresponding rows of $X$ and columns of $A$, respectively.  The
following noiseless version of the canonical MMV formulation is very useful for our analysis.
\begin{definition}[Canonical form noiseless MMV]\label{def:can}
Let $m$, $n$ and $r$ be positive integers ($r\leq m<n$) that
represent the number of sensor elements, the ambient space
dimension, and the number of snapshots, respectively. Suppose that
we are given a sensing matrix $A\in\mathbb{R}^{m\times n}$ and an
observation matrix $B\in \mathbb{R}^{m\times r}$ such that
$B=AX_{*}$ for some $X_{*}\in\mathbb{R}^{n\times r}$ and
$\|X_{*}\|=|{\rm supp}X|=k$. A canonical form noiseless multiple
measurement vector (MMV) problem is given the estimation problem of
$k$-sparse vectors $X\in\mathbb{R}^{n\times r}$ through multiple
snapshots $B=AX$ using the following formulation:
\begin{eqnarray}\label{eqdefcan_mmv}
{\rm minimize}~~~\|X\|_0\\
{\rm subject~to}~~~B=AX, \notag
\end{eqnarray}
where $\|X\|_0=|{\rm supp}X|$, ${\rm supp}X=\{1\leq i\leq n :
\mathbf{x}^i\neq 0\}$, $\mathbf{x}^i$ is the $i$-th row of $X$, and
the observation matrix $B$ is full rank, i.e. ${\rm rank}(B)=r\leq
k$.
\end{definition}

Compared to \eqref{eqdefmmv}, the canonical form MMV has the additional
constraint that ${\rm rank}(B)=r \leq \|X\|_0$. This is not
problematic though since every MMV problem  can be converted into
a canonical form using the following dimension reduction.
\begin{itemize}
   \item Suppose we are given the following linear sensor observations: $B=AX$ where  $A\in\mathbb{R}^{m\times n}$ and
   $X\in\mathbb{R}^{n\times l}$ satisfies $\|X\|_0=k$.
   \item Compute the SVD as $B=UD_rV^{*}$, where $D_r$ is an $r\times r$ diagonal matrix, $V\in\mathbb{C}^{l\times r}$ consists of right singular vectors,
    and $r={\rm rank}(B)$, respectively.
   \item Reduce the dimension as $B_{SV}=BV$ and $X_{SV}=XV$.
   \item The resulting canonical form MMV becomes $B_{SV}=AX_{SV}$.
\end{itemize}
We can easily show that  ${\rm rank}(B_{SV})=r \leq k$  and
 the sparsity $k:=\|X\|_0 = \|X_{SV}\|_0$ with probability 1.
Therefore, without loss of generality,  the canonical form of the MMV in Definition~\ref{def:can}
is assumed throughout the paper.

The following definitions are  used throughout this paper.
\begin{definition}\cite{Donoho2005pnas}
The rows (or columns) in $\mathbb{R}^n$  are in general position if any $n$ collection of rows (or columns) are linearly independent.
\end{definition}

If $A\in\mathbb{R}^{m\times n}$, where $m<n$, the columns of $A$ are in general
position if and only if ${\rm spark}(A)=m+1$. Also, it is equivalent to $K$-${\rm rank}(A)=m$ where
$K-$rank denotes the Kruscal rank, where a Kruscal rank of $A$ is the maximal number $q$ such that every collection of $q$ columns of $A$ is linearly independent \cite{Mishali08rembo}.
\begin{definition}[Mutual coherence]
For a sensing matrix $A=[\ab_1,\cdots,\ab_n]\in\mathbb{R}^{m\times
n}$, the mutual coherence $\mu(A)$ is given by
$$\mu=\max\limits_{1\leq j< k\leq n}\frac{|\ab_j^{*}\ab_k|}{\|\ab_j\|\|\ab_k\|},$$ where the
superscript $^*$ denotes the Hermitian transpose.
\end{definition}

\begin{definition}[Restricted Isometry Property (RIP)]
A sensing matrix $A\in\mathbb{R}^{m\times n}$ is said to have a
$k$-restricted isometry property (RIP) if there exist left and right
RIP constants $0\leq \delta^L_k, \delta^R_k<1$  such that
$$(1-\delta^L_k)\|\mathbf{x}\|^2\leq \|A\mathbf{x}\|^2\leq (1+\delta^R_k)\|\mathbf{x}\|^2$$
for all $\mathbf{x}\in\mathbb{R}^n$ such that $\|\mathbf{x}\|_0\leq
k$. A single RIP constant $\delta_k = \max\{\delta^L_k,
\delta^R_k\}$ is often referred to as the RIP constant.
\end{definition}

\bigskip
Note that the condition for the left RIP constant $0\leq \delta^L_{2k}<1$  is
sufficient for the uniqueness of any $k$-sparse vector $\xb$ satisfying $A\xb=\bb$ for any
$k$-sparse vector $\xb$, but the condition $\delta_{2k}<1$ is often too
restrictive.


%

\section{Conventional MMV  Algorithms }\label{sec:review}
\setcounter{equation}{0} \setcounter{theorem}{0} \indent

In this section, we review the conventional algorithms for the MMV
problem and analyze their limitations. This survey is useful in order
to understand the necessity of developing a new class of algorithms.
Except for the MUSIC and cumulant MUSIC algorithm, all other algorithms have been
developed in the context of compressive sensing. We will show that all the existing methods have their own disadvantages. In particular, the maximum sparsity levels that can be resolved by these algorithms are limited in achieving the maximum gain from joint sparse recovery.

\subsection{Simultaneous Orthogonal Matching Pursuit (S-OMP)\cite{tropp2006ass,chen2006trs}}
\label{sssec:subsubhead} The S-OMP algorithm is a greedy algorithm that performs the following procedure:
\begin{itemize}
\item at the first iteration, set $B_0=B$ and $S_0=\emptyset$,
\item after $J$ iterations, $S_j=\{l_j\}_{j=1}^J$ and $B_J=(I-P_{S_J})B$, where $P_{S_J}$ is the orthogonal projection onto ${\rm span}\{\mathbf{a}_{l_j}\}_{j=1}^J$,
\item select $l_{J+1}$ such that
$\|\mathbf{a}_{l_{J+1}}^{*}B_J\|_2=\max\limits_{1\leq l\leq
N}\|\mathbf{a}_l^{*}B_J\|_2$ and set $S_{J+1}=S_{J}\cup
\{l_{J+1}\}.$
\end{itemize}
Worst case analysis of S-OMP \cite{gribonval2008atoms} shows that
a sufficient condition for S-OMP to succeed is
\begin{eqnarray}\label{tropp-erc}
\max\limits_{j\in {\rm supp}X}\|A_S^{\dagger}\ab_j\|_1<1,
\end{eqnarray}
where $S={\rm supp}X$. An explicit form of  recoverable sparsity
level  is then given by
\begin{eqnarray}\label{l0bound-mc}
  \|X\|_0 &< & \frac{1}{2}\left( \frac{1}{\mu} +  1 \right) \ .
\end{eqnarray}
Note that these conditions are exactly the same as Tropp's exact
recovery conditions for the SMV problem \cite{tropp2004gg}, implying
that the sufficient condition for the maximum sparsity level is not improved with an increasing
number of snapshots even in the noiseless case. In order to resolve
this issue, the authors in \cite{gribonval2008atoms} and
\cite{eldar2010average} performed an average case analysis for S-OMP, and
showed that S-OMP can recover the input signals for the MMV problem with
higher probability when the number of snapshots increases. However,
the simulation results in \cite{gribonval2008atoms} and
\cite{eldar2010average} suggest that S-OMP performance is saturated after
some number of snapshots, even with noiseless measurements, and S-OMP
never achieves the $l_0$ bound with a finite number of snapshots.

\subsection{2-Thresholding \cite{gribonval2008atoms}}
 In 2-thresholding, we select a set $S$ with
$|S|=k$ such that
$$\|\mathbf{a}_l^{*}B\|_2\geq \|\mathbf{a}_j^{*}B\|_2,~~{\rm for~all}~l\in S,~j\notin S.$$
If we estimate the ${\rm supp}X$ by the above criterion, we can recover
the nonzero component of $X$ by the equation $X^S=A_S^{\dagger}Y$.
In \cite{gribonval2008atoms}, the authors demonstrated that the
performance of 2-thresholding is often not as good as that of S-OMP,
which suggests that 2-thresholding never achieves the $l_0$-bound
\eqref{l0-bound-mmv} with finite snapshots even if the measurements are
noiseless.
%

\subsection{ReMBO algorithm \cite{Mishali08rembo}}

Reduce MMV and Boost (ReMBo) by Mishali and Eldar
\cite{Mishali08rembo} addresses the MMV problem by reducing it to a series of
SMV problems based on the following.
\begin{theorem}\cite{Mishali08rembo}\label{th-rembo}
Suppose that $X$ satisfies $\|X\|_0=k$ and $AX=B$ with $k<{\rm
spark}(A)/2$. Let $\mathbf{v}\in \mathbb{R}^r$ be a random vector
with an absolutely continuous distribution and define
$\mathbf{b}=A\mathbf{v}$ and $\overline{\mathbf{x}}=X\mathbf{v}$.
Then, for a random SMV system $A\mathbf{x}=\mathbf{b}$ and $\bb=B\vb$, we have
\begin{itemize}
\item [(a)] For every $\mathbf{v}$, the vector $\overline{\mathbf{x}}$ is the unique $k$-sparse solution.
\item [(b)] ${\rm Prob}({\rm supp}(\xb)={\rm supp}(\overline{\mathbf{x}}))=1.$
\end{itemize}
\end{theorem}

Employing the above theorem, Mishali and Eldar
\cite{Mishali08rembo} proposed the ReMBo algorithm which performs the following procedure:
\begin{itemize}
   \item set the maximum number of iterations as $\textsf{MaxIters}$, set $i=1$ and $\textsf{Flag}=\textsf{F}$,
   \item while $i\leq \textsf{MaxIters}$ and $\textsf{Flag}=\textsf{F}$, generate a random SMV problem as in Theorem \ref{th-rembo},
       \begin{itemize}
          \item if the SMV problem has a $k$-sparse solution, then we let $S$ be the support of the solution vector, and let $\textsf{Flag}=\textsf{T}$
          \item otherwise, increase $i$ by 1
       \end{itemize}
   \item if $\textsf{Flag}=\textsf{T}$, find the nonzero components of $X$ by the equation $X^S=A_S^{\dag}B$.
\end{itemize}
In order to achieve the $l_0$ bound \eqref{l0-bound-music}  by ReMBO without any combinatorial SMV solver,
an uncountable number of random vectors $\vb$  are required. With a finite
number of choices of $\vb$, the performance of ReMBo is therefore
dependent on randomly chosen input and the solvability of a randomly
generated SMV problem so that it is difficult to achieve the
theoretical $l^0$-bound even with noiseless measurements.

\subsection{Mixed norm approach \cite{tropp2006algorithms}}
The mixed norm approach is an extension of the convex relaxation method
in SMV \cite{Tr06} to the MMV problem. Rather than solving the original
MMV problem \eqref{eqdefcan_mmv}, the mixed norm approaches solve
the following convex optimization problem:
\begin{eqnarray}
{\rm minimize}~~~\|X\|_{p,q} & \quad 1\leq p, q \leq 2\\
{\rm subject~to}~~~B=AX, \notag &
\end{eqnarray}
where $\|X\|_{p,q}=(\sum_{i=1}^n \|\xb^i\|_p^q)^{\frac{1}{q}}$. The
optimization problem can be formulated as an SOCP (second order cone
program) \cite{malioutov2005ssr}, homotopy continuation
\cite{bach2008consistency}, and so on. Worst case bounds for the mixed
norm approach were derived in \cite{chen2006trs}, which shows no
improvement with the increasing number of measurement. Instead, Eldar {\em et
al} \cite{eldar2010average} considered the average case analysis when $p=2$
and $q=1$ and showed that if
$$\max\limits_{j\notin S}\|A_S^{\dagger}\ab_j\|_2\leq \alpha<1,$$
where $S={\rm supp}X$, then the probability success recovery of
joint sparsity increases with the number of snapshots. However,
it is not clear whether this convex relaxtion can achieve the $l_0$
bound.

\subsection{Block sparsity approaches \cite{eldar2009compressed}}
Block sparse signals have been extensively studied by Eldar et
al  using the uncertainty relation for the block-sparse signal and block
coherence concept. Eldar et al. \cite{eldar2009compressed} showed
that the block sparse signal can be efficiently recovered  using a fewer
number of measurements by exploiting the block sparsity pattern as
described in the following theorem:
%

\begin{theorem}\cite{eldar2009compressed}
Let positive integers $L,n,N$ and $D=[\Db[1],\cdots,\Db[n]]\in\mathbb{R}^{L\times N}$ be given, where $L<N$, $N=nr$ for some positive integer $r$ and for each $1\leq j\leq n$, $\Db[j]\in\mathbb{R}^{L \times r}$. Let $\mu_B$ be the block-coherence which is defined by
$$\mu_B=\max\limits_{1\leq j<k\leq n}\frac{1}{r}\rho(\Db[j]^{*}\Db[k])$$
where $\rho$ denotes the spectral radius, $\nu$ be the sub-coherence of the
sensing matrix $A$ which is defined by
$$\nu=\max\limits_l\max\limits_{i\neq j}|\db_i^{*}\db_j|,~~\db_i,\db_j\in \Db[l],$$
and $r$ be the block size. Then, the block OMP and
block mixed $l_2/l_1$ optimization program  successfully recover the
$k$-block sparse signal if
\begin{equation}\label{eldar-l0}
    k r < \frac{1}{2} \left(\mu_B^{-1} + r - (r-1) \frac{\nu}{\mu_B}
    \right).
\end{equation}
\end{theorem}

\bigskip

Note that we can transform $B=AX$ into an SMV system ${\rm
vec}(B^T)=(A\bigotimes I_r){\rm vec}(X^T)$, where ${\rm vec}(X^T)$
is block-$k$ sparse with length $r$ and $\bigotimes$ denotes the
Kronecker product of matrices. Therefore, one may think that  we can
use the block OMP or block $l_2/l_1$ optimization problem to solve
the MMV problem. However, the following theorem shows that this is
pessimistic.
\begin{theorem}
For the canonical MMV problem in Definition~\ref{def:can}, a sufficient condition 
for recovery using block-sparsity  is 
\begin{equation}\label{eldar-l0-2}
    k < \frac{1}{2} \left(\mu^{-1} + 1     \right),
\end{equation}
where $\mu$ denotes the mutual coherence of the sensing matrix $A
\in \Rd^{m\times n}$.
\end{theorem}
\begin{proof}
Since $A\bigotimes I_r=[a_{i,j}I_r]_{i,j=1}^{m,n}$, if we let
$A\bigotimes I_r=[\Db[1],\cdots,\Db[n]]$, we have $\nu=0$ due to the diagonality,
and
$$\mu_B=\max\limits_{1\leq j<k\leq n}\frac{1}{r}\rho\left(\Db[j]^{*}\Db[k]\right)=
\max\limits_{1\leq j<k\leq
n}\frac{1}{r}\rho\left(\sum\limits_{i=1}^m
a_{ij}^{*}a_{ik}I\right)=\frac{\mu}{r}$$ by the definition of mutual
coherence. Applying (\ref{eldar-l0}) with $\nu=0$ and $\mu_B=\mu/r$,
we obtain (\ref{eldar-l0-2}).
\end{proof}

Note that \eqref{eldar-l0-2} is the same as that of OMP for SMV. The
main reason for the failure of the block sparse approach for the MMV problem
is that the block sparsity model does not exploit the diversity of
unknown matrix $X$. For example, the block sparse model cannot
differentiate a rank-one input matrix $X$ and full-rank matrix $X$.

\subsection{M-SBL \cite{wipf2006bayesian}}

M-SBL (Sparse Bayesian Learning) by Wipf and Rao \cite{wipf2007ebs}
is a Bayesian compressive sensing algorithm to address the $l_0$
minimization problem.  M-SBL is based on the ARD (automatic relevance
determination) and utilizes an empirical Bayesian prior thereby
enforcing a joint sparsity.
%
%
%
Specifically, the M-SBL performs the following procedure:
\begin{itemize}
\item [(a)] initialize $\mathbf{\gamma}$ and $\Gamma:={\rm diag}(\mathbf{\gamma}) \in \Rd^{n\times n}$.
\item [(b)] compute the posterior variance $\Sigma$ and mean $\hat{X}$
as follows:
\begin{eqnarray*}\label{eq-msbl-moments}
\Sigma&:=&\Gamma-\Gamma A^*(A\Gamma A^{*}+\lambda I)^{-1}A\Gamma\\
\hat X&:=&\Gamma A^{*}(A\Gamma A^{*}+\lambda I)^{-1}B,
\end{eqnarray*}
where $\lambda>0$ denotes a regularization parameter.
\item [(c)] update $\mathbf{\gamma}$ by
\begin{eqnarray*}
  \gamma_j^{(new)} &=&
  \frac{\|\mathbf{\mu}_j\|^2}{r}\frac{1}{1-\gamma_j^{-1}\Sigma_{jj}},
  \quad 1\leq j \leq n \,
\end{eqnarray*}
\item [(d)] repeat (b) and (c) until $\mathbf{\gamma}$ converges to some fixed point
$\mathbf{\gamma}^{*}$.
\end{itemize}
Wipf and Rao \cite{wipf2007ebs} showed that increasing the number of
snapshots in SBL reduces the number of local minimizers so that
the possibility of recovering input signals increases from  joint
sparsity. Furthermore, in the noiseless setting, if we have $k$
linearly independent measurements and the nonzero rows of $X$ are
{\em orthogonal},  there is a unique fixed point
$\mathbf{\gamma}^{*}$ so that we can correctly recover the
$k$-sparse input vectors. To the best of our knowledge,  M-SBL is the only
compressive sensing algorithm that achieves the same $l_0$-bound as
MUSIC when $r=k$. However, the orthogonality condition for the input
vector $X$ that achieves the maximal sparsity level is more
restricted than that of MUSIC. Furthermore, no explicit
expression for the maximum sparsity level was provided for
the range ${\rm rank}(B)< k$.

%
%
%
%

\subsection{The MUSIC Algorithm \cite{Feng97,schmidt1986multiple}}
The MUSIC algorithm was originally developed to estimate the
continuous parameters such as bearing angle or DOA.  However, the
MUSIC criterion can be still modified to identify the support set
from the finite index set as follows.
\begin{theorem}\cite{Feng97,schmidt1986multiple}(MUSIC Criterion)
Assume that we have $r$ linearly independent measurements
$B\in\mathbb{R}^{m\times r}$
such that $B=AX_*$  for $X_{*}\in\mathbb{R}^{n\times r}$ and $r=\|X_{*}\|_0=:k<m$. Also, we assume that
the columns of a sensing matrix $A\in\mathbb{R}^{m\times n}$ are in
general position; that is, any collection of $m$ columns of $A$ are
linearly independent. Then, for any $j\in \{1,\cdots,n\}$, $j\in {\rm
supp}X_{*}$ if and only if
\begin{equation}\label{music-cond}
Q^{*}\mathbf{a}_j=0,
\end{equation}
or equivalently
\begin{equation}\label{music-cond2}
\ab_j^{*}P_{R(Q)}\ab_j=0
\end{equation}
where $Q\in\mathbb{R}^{m\times (m-r)}$  consists of orthonormal
columns such that $Q^{*}B=0$ so that $R(Q)^{\perp}=R(B)$, which is
often called ``noise subspace''. Here, for matrix $A$, $R(A)$ denotes the range space of $A$.
\end{theorem}
\begin{proof}
By the assumption, the matrix of multiple measurements $B$ can be factored as a product $B=A_SX_{*}^S$ where $A_S\in\mathbb{R}^{m\times k}$ and $X_{*}^S\in\mathbb{R}^{k\times k}$, where $S={\rm supp}X_{*}$, $A_S$ is the matrix
 which consists of columns whose indices are in $S$ and $X_{*}^S$ is the matrix that consists of rows whose indices are in $S$. Since $A_S$ has full column rank and $X_{*}^S$ has full row rank, $R(B)=R(A_S)$. Then we can obtain a singular value decomposition as
$$B=[U~Q]{\rm diag}[\sigma_1,\cdots,\sigma_k,0,\cdots,0]V^*,$$
where $R(U)=R(A_S)=R(Q)^{\perp}$.
Then, $Q^{*}\mathbf{a}_j=0$ if and only if $\mathbf{a}_j\in R(Q)^{\perp}=R(A_S)$ so that $\mathbf{a}_j$ can be expressed as a linear combination of $\{\mathbf{a}_k\}_{k\in S}$. Since the columns of $A$ are in general position, $Q^{*}\mathbf{a}_j=0$ if and only if $j\in {\rm supp}X_{*}$.
\end{proof}

Note that the MUSIC criterion \eqref{music-cond} holds for all $m\geq
k+1$ if the columns of $A$ are in general position. Using the
compressive sensing terminology, this implies that the recoverable
sparsity level by MUSIC (with a probability 1 for the noiseless
measurement case) is given by
\begin{equation}\label{eq:max_music}
    \|X\|_0 < m = {\rm spark}(A)-1,
\end{equation}
where the last equality comes from the definition of the ${\rm spark}$.
Therefore, the $l_0$ bound \eqref{l0-bound-mmv} can be achieved by
MUSIC when $r=k$. However, for any $r<k$, the MUSIC condition
\eqref{music-cond} does not hold. This  is a major drawback of MUSIC
compared to the compressive sensing algorithms that allows perfect
reconstruction with extremely large probability by increasing the
number of sensor elements, $m$.

\subsection{Cumulant MUSIC}

The fourth-order cumulant or higher order MUSIC was proposed by
Porat and Friedlander \cite{Porat1991HighOrderMusic} and Cardoso
\cite{cardoso1989source} to improve the number of resolvable
resolvable sources over the conventional second-order MUSIC.
Specifically, the cumulant MUSIC derives a MUSIC-type subspace
criterion from the cumulant of the observation matrix. It has been
shown that the cumulant MUSIC can resolve  more sources than
conventional MUSIC for specific array geometries
\cite{gonen1997applications}. However, a significant increase in the
variance of the target estimate of a weak source in the presence of
stronger sources has been reported, which was not observed for
second order MUSIC \cite{cardoso1995asymptotic}.
 This increase often  prohibit the use
of fourth-order methods, even for large SNR, when the dynamic
range of the sources is important \cite{cardoso1995asymptotic}.
Furthermore,  for general array geometries, the performance of the cumulant MUSIC is not clear.
Therefore, we need to develop a new type of algorithm that can overcome these drawbacks.

\subsection{Main Contributions of Compressive MUSIC}

Note that the existing MMV compressive sensing approaches are  based
on a probabilistic guarantee, whereas  array signal processing
provides a deterministic guarantee. Rather than taking such extreme
view points to address a MMV problem, the main contribution of
CS-MUSIC is to show that we should take the best of both approaches.
More specifically, we  show that as long as $k-\mathrm{rank}(B)$
partial support can be estimated with any compressive sensing
algorithms, the remaining unknown support of $\mathrm{rank}(B)$ can
be estimated deterministically using a novel generalized MUSIC
criterion. By allowing such hybridization,  our CS-MUSIC can
overcome the drawbacks of the all existing approaches and achieves
the superior recovery performance that had not been achievable by
any of the aforementioned MMV algorithms. Hence, the following
sections discuss what conditions are required for the generalized
MUSIC and partial support support recovery to succeed, and how
CS-MUSIC outperforms existing methods.

\section{Generalized MUSIC criterion for compressive MUSIC}\label{sec:gmusic}
\label{sec:model}

\setcounter{equation}{0} \setcounter{theorem}{0} \indent

This section derives an important component of compressive MUSIC,
which we call the generalized MUSIC criterion. This extends the
MUSIC criterion \eqref{music-cond} for $r\leq k$. Recall that when
we obtain $k$ linearly independent measurement vectors,  we can
determine the support of multiple signals with the condition that
$Q^*\mathbf{a}_j=0$ if and only if $j\in {\rm supp}X$. In general,
if we have $r$ linearly independent measurement vectors, where
$r\leq k$, we have the following.
\begin{theorem}[Generalized MUSIC criterion]\label{lemspark}
  Let $m$, $n$ and $r$ be positive integers such that $r\leq m<n$. Suppose that we are given a sensing matrix
  $A\in\mathbb{R}^{m\times n}$ and an observation matrix $B\in\mathbb{R}^{m\times r}$.
   Assume that the MMV problem  is in canonical form, that is,  ${\rm rank}(B)=r\leq k$. Then, the following holds:\\
   (a) ${\rm spark}(Q^*A)\leq k-r+1.$\\
   (b) If the $k$ nonzero rows are in general position (i.e., any collection of $r$ nonzero rows are linearly independent) and $A$ satisfies the RIP condition with $0\leq \delta^L_{2k-r+1}(A)<1$, then
   $${\rm spark}(Q^*A)= k-r+1.$$
\end{theorem}
\begin{proof}
See Appendix A.
\end{proof}

Note that, unlike the classical MUSIC criterion, a condition for the
left RIP constant $0\leq \delta^L_{2k-r+1}(A)<1$  is required in
Theorem~\ref{lemspark} (b). This condition has the following very interesting
implication.

\begin{lemma}
For the canonical form MMV, $A\in\mathbb{R}^{m\times n}$
satisfies RIP with $0\leq \delta^L_{2k-r+1}<1$ if and only if
\begin{equation}\label{eq-k<l0}
k<\frac{{\rm spark}(A)+{\rm rank}(B)-1}{2}.
\end{equation}
\end{lemma}
\begin{proof}
Since $A\in\mathbb{R}^{m\times n}$  has the left RIP condition
$0\leq \delta^L_{2k-r+1}<1$, any collection of $2k-r+1$ columns of $A$
are linearly independent so that ${\rm spark}(A)>2k-r+1$. Hence,
$$k<\frac{{\rm spark}(A)+r-1}{2}=\frac{{\rm spark}(A)+{\rm
rank}(B)-1}{2}$$ since $r={\rm rank}(B)$. For the converse, assume the condition \eqref{eq-k<l0}. Then we have $2k-r+1<{\rm spark}(A)$ which implies $0\leq\delta_{2k-r+1}^L<1$.
\end{proof}

Hence, if $A$ satisfies RIP with $0\leq \delta_{2k-r+1}^L<1$ and if
we have $k$-sparse coefficient matrix $X$ that satisfies $AX=B$,
then $X$ is the unique solution of the MMV. In other words, under
the above RIP assumption, for noiseless case we can achieve the
$l_0$-uniqueness bound, which is the same as the theoretical limit
\eqref{l0-bound-mmv}. Note that when $k=r$, we have ${\rm
spark}(Q^{*}A)=1$, which is equivalent to there being some $j$'s
such that $Q^{*}\mathbf{a}_j=0$, which is equivalent to the
classical MUSIC criterion.
 By the above lemma, we can obtain a
{\em generalized MUSIC criterion} for the case $r\leq k$ in the
following theorem.
\begin{theorem}\label{com-music}
Assume that $A\in\mathbb{R}^{m\times n}$, $X\in\mathbb{R}^{n\times
r}$, and $B\in\mathbb{R}^{m\times r}$ satisfy $AX=B$ and the
conditions in Theorem \ref{lemspark} (b). If $I_{k-r}\subset {\rm supp}X$
with $|I_{k-r}|=k-r$ and $A_{I_{k-r}}\in \mathbb{R}^{m\times(k-r)}$,
which consists of columns, whose indices are in $I_{k-r}$. Then for
any $j\in \{1,\cdots,n\}\setminus I_{k-r}$,
\begin{equation}\label{eq-comusic}
{\rm rank}(Q^*[A_{I_{k-r}},\mathbf{a}_j])=k-r
\end{equation}
if and only if $j\in {\rm supp}X$.
\end{theorem}
\begin{proof}
See Appendix B.
\end{proof}

When $r=k$,  $A_{I_{k-r}}=\emptyset$ and \eqref{eq-comusic} is the
same as the classic MUSIC criterion \eqref{music-cond} since  ${\rm
rank}(Q^*\ab_j)= 0 \Longleftrightarrow Q^*\ab_j=0$. However, the
generalized MUSIC criterion \eqref{eq-comusic} for $r<k$ is based on
the rank of the matrix, which is prone to error under an incorrect estimate
of noise subspace $Q$ when the measurements are corrupted by additive
noise. Hence, rather than using \eqref{eq-comusic}, the following
equivalent criterion is more practical.
\begin{corollary}\label{coro-comusic}
Assume that $A\in\mathbb{R}^{m\times n}$, $X\in\mathbb{R}^{n\times
r}$, $B\in\mathbb{R}^{m\times r}$, $I_{k-r}\subset {\rm supp}X$,
and $A_{I_{k-r}}$ are the same as in Theorem \ref{com-music}. Then,
\begin{equation}\label{eq-comusicmod}
\ab_j^{*}\left[P_{R(Q)}-P_{R(P_{R(Q)}A_{I_{k-r}})}\right]\ab_j=0
\end{equation}
if and only if $j\in {\rm supp}X$.
\end{corollary}
\begin{proof}
See Appendix B.
\end{proof}

Note that $P_{R(Q)}=QQ^{*}$ in MUSIC criterion \eqref{music-cond2} is now replaced by $P_{R(Q)}-P_{R[P_{R(Q)}A_{I_{k-r}}]}$ where $I_{k-r}\subset {\rm supp}X$. The following theorem shows that $P_{R(Q)}-P_{R[P_{R(Q)}A_{I_{k-r}}]}$ has very important geometrical meaning.

\begin{theorem}\label{cmusic-geo}
Assume that we are given a noiseless MMV problem which is in
canonical form. Also, suppose that $A$ and $X$ satisfy the
conditions as in Theorem \ref{lemspark} (b). Let
$U\in\mathbb{R}^{m\times r}$and $Q\in\mathbb{R}^{m\times (m-r)}$
consist of orthonormal columns such that $R(U)=R(B)$  and
$R(Q)^{\perp}=R(B)$. Then the following properties hold :
\begin{itemize}
\item [(a)] $UU^{*}+P_{R(QQ^{*}A_{I_{k-r}})}$ is equal to the orthogonal projection onto $R(B)+R(QQ^{*}A_{I_{k-r}})$.
\item [(b)] $QQ^{*}-P_{R(QQ^{*}A_{I_{k-r}})}$ is equal to the orthogonal projection onto $R(Q)\cap R(QQ^{*}A_{I_{k-r}})^{\perp}$.
\item [(c)] $QQ^{*}-P_{R(QQ^{*}A_{I_{k-r}})}$ is equal to the orthogonal complement of $R([U~A_{I_{k-r}}])$ or $R([B~A_{I_{k-r}}])$.
\end{itemize}
\end{theorem}
\begin{proof}
See Appendix C.
\end{proof}

\begin{figure}[htbp]
 \centerline{\epsfig{figure=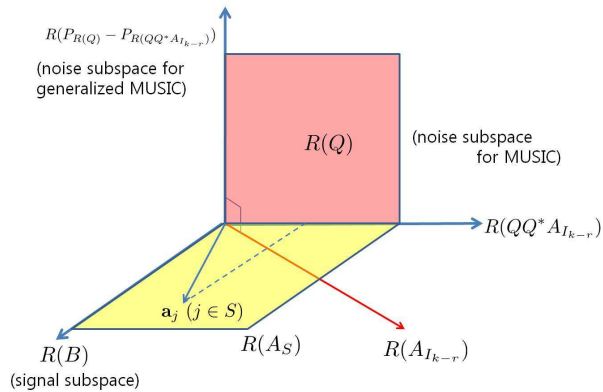,width=8cm}}
  \caption{Geometric view for the generalized MUSIC criterion : the dashed line corresponds to the conventional MUSIC criterion, where the squared norm of the projection of $\ab_j(j\in {\rm supp}X)$ onto the noise subspace $R(Q)$ may not be zero.  $\ab_j(j\in {\rm supp}X)$ is orthogonal to the subspace
$R(P_{R(Q)}-P_{R(QQ^{*}A_{I_{k-r}})})$ so that we can identify the indices of the support of $X$ with the generalized MUSIC criterion.}
 \label{fig:cmusic-geo}
\end{figure}

Figure \ref{fig:cmusic-geo} illustrates the geometry of corresponding subspaces. Unlike the MUSIC, the orthogonality of the $\ab_j$, $j\in{\rm supp}X$ need to be checked with respect to $R(Q)\cap R(QQ^{*}A_{I_{k-r}})^{\perp}$. Based on the geometry, we can obtain following algorithms for support detection.

{\bf (Algorithm 1: Original form)}

\begin{enumerate}
   \item Find $k-r$ indices of ${\rm supp}X$ by any MMV compressive sensing algorithms such as 2-thresholding or SOMP.
   \item Let $I_{k-r}$ be the set of indices which are taken in Step 1 and $S=I_{k-r}$.
   \item For $j\in \{1,\cdots,n\}\setminus I_{k-r}$,  calculate the quantities $\eta(j)=\ab_j^{*}[P_{R(Q)}-P_{P_{R(Q)}A_{I_{k-r}}}]\ab_j$ for all $j\notin
   I_{k-r}$.
   \item Make an ascending ordering of $\eta(j)$, $j\notin
   I_{k-r}$, choose indices that correspond to the first $r$
   elements, and put these indices into $S$.
\end{enumerate}

{\bf (Algorithm 2: Signal subspace form)}

Alternatively, we can also use the signal subspace form to identify the support of $X$:
\begin{enumerate}
   \item Find $k-r$ indices of ${\rm supp}X$ by any MMV compressive sensing algorithms such as 2-thresholding or SOMP.
   \item Let $I_{k-r}$ be the set of indices which are taken in Step 1 and $S=I_{k-r}$.
   \item For $j\in \{1,\cdots,n\}\setminus I_{k-r}$,  calculate the quantities $\eta(j)=\ab_j^{*}[P_{R(U)}+P_{R(P_{R(U)}^{\perp}A_{I_{k-r}})}]\ab_j$ for all $j\notin
   I_{k-r}$.
   \item Make a descending ordering of $\eta(j)$, $j\notin
   I_{k-r}$, choose indices that correspond to the first $r$
   elements, and put these indices into $S$.
\end{enumerate}

\bigskip

In compressive MUSIC, we determine $k-r$ indices of ${\rm supp}X$
with CS-based algorithms such as 2-thresholding or S-OMP, where the
exact reconstruction is a probabilistic matter. After that process,
we recover remaining $r$ indices of ${\rm supp}X$ with a generalized
MUSIC criterion, which is given in Theorem \ref{com-music} or
Corollary~\ref{coro-comusic}, and this reconstruction process is
deterministic. This hybridization makes the compressive MUSIC
applicable for all ranges of $r$, outperforming all the existing
methods.

So far, we have discussed about the recovery of the support of the
multiple input vectors assuming that we know about the size of the
support.
One of the disadvantages of the existing MUSIC-type algorithms is that if the sparsity level is overestimated, spurious peaks are often observed.
However, in CS-MUSIC  when we do not know about the correct size of the support,
we can still apply the following lemma to estimate the size of the
support.

\begin{lemma}\label{lem-supp-est}
Assume that $A\in\mathbb{R}^{m\times n}$, $X_{*}\in\mathbb{R}^{n\times
r}$ and $B\in\mathbb{R}^{m\times r}$ satisfy $AX_{*}=B$ and the
conditions in theorem \ref{lemspark} (b), and $k$ denotes the true
sparsity level, i.e. $k=\|X_{*}\|_0$. Also, assume that $r< {\hat k}\leq
k+r$ and we are given $I_{{\hat k}-r}\subset {\rm supp}X$ with
$|I_{{\hat k}-r}|={\hat k}-r$, where $I_{{\hat k}-r}$ is the partial support of size ${\hat k}-r$ estimated by any MMV compressive sensing algorithm. Also, we let
$\eta(j):=\ab_j^{*}[P_{R(Q)}-P_{R(P_{R(Q)}A_{I_{k-r}})}]\ab_j$. Then,
${\hat k}=k=\|X_{*}\|_0$ if and only if
\begin{equation}\label{eq-supp-est}
C({\hat k}):=\min\limits_{J\cap I_{{\hat k}-r}=\emptyset,
|J|=r}\sum\limits_{j\in J}\eta(j)=0.
\end{equation}
\end{lemma}
\begin{proof}
Necessity is trivial by Corollary \ref{coro-comusic} so we only need to show sufficiency of (\ref{eq-supp-est}) assuming the contrary. We divide the proof into two parts. \\
(i) $r<{\hat k}<k$ : By the Lemma \ref{lemspark}, for any $j\in
\{1,\cdots,n\}\setminus I_{{\hat k}-r}$,
$${\rm rank}[Q_{I_{{\hat k}-r}}^*[A_{{\hat k}-r},\ab_j]]={\hat k}-r+1.$$
As in the proof of Corollary \ref{coro-comusic}, this implies $\eta(j)>0$ for any $j\in\{1,\cdots,n\}\setminus I_{{\hat k}-r}$, so that we have $C({\hat k})> 0$ for ${\hat k}<k$.\\
(ii) $k<{\hat k}\leq k+r$ : Here, we have already chosen  at least
$k-r+1$ indices of the support of $X$. By Corollary
\ref{coro-comusic}, (\ref{eq-comusicmod}) holds only for, at most,
$r-1$ elements of $\{1,\cdots,n\}\setminus I_{{\hat k}-r}$ since
$I_{{\hat k}-r}\subset {\rm supp}X$. Hence, $C({\hat k})>0$ for ${\hat k}>k$.
\end{proof}
The minimization in \eqref{eq-supp-est} is over all index sets $J$ of size $r$ that include elements from $\{1,\cdots,n\}$ and no elements form $I_{{\hat k}-r}$. For fixed ${\hat k}$ and $I_{{\hat k}-r}$, this minimization can be performed by first computing the summands for all
$j\in \{1,\cdots,n\}\setminus I_{{\hat k}-r}$ and then selecting the $r$ of smallest magnitude.  Lemma \ref{lem-supp-est} also tells us that if we calculate $C({\hat k})$ by
increasing ${\hat k}$ from $r$,  then  the first  ${\hat k}$ such that
 $C({\hat k})=0$ corresponds to the unknown sparsity level. For noisy measurements, we can choose the first local minimizer of $C({\hat k})$ by increasing ${\hat k}$.


\bigskip

\section{Sufficient Conditions for Sparse Recovery using Compressive MUSIC}\label{sec:no}
\setcounter{equation}{0} \setcounter{theorem}{0} \indent

\subsection{Large system MMV model}
Note that the recovery performance of compressive MUSIC relies
entirely on the correct identification of $k-r$  partial support in ${\rm
supp}X$ via compressive sensing approaches  and the remaining  $r$
indices using the generalized MUSIC criterion. In practice, the measurements are noisy, so the theory we derived for noiseless measurement should be modified. In this section, we
derive sufficient conditions for the minimum number of sensor
elements (the number of rows in each measurement vector) that
guarantee the correct support recovery by compressive MUSIC. Note that for the success of compressive MUSIC, both CS step and the generalized MUSIC step should succeed. Hence, this section derives separate conditions for each step, which is required for the success of compressive MUSIC.

For  SMV compressive sensing, Fletcher, Rangan and Goyal
\cite{fletcher2009necessary} derived an explicit expression for the
minimum number of sensor elements for the 2-thresholding algorithm
to find the correct support set. Also, Fletcher and Rangan
\cite{fletcher-orthogonal} derived a sufficient condition for S-OMP to recover $X$.
Even though their derivation is based on a large system model with
a Gaussian sensing matrix, it has provided very useful insight into
the SMV compressive sensing problem. Therefore, we employed a
large system model to derive a sufficient condition for compressive
MUSIC.
\begin{definition}
A large system noisy canonical MMV model,
$\mathrm{LSMMV}(m,n,k,r;\epsilon)$, is defined as an estimation
problem of $k$-sparse vectors $X\in\mathbb{R}^{n\times r}$ that
shares a common sparsity pattern through multiple noisy snapshots
$Y=AX+N$ using the following formulation:
\begin{eqnarray}\label{mmv-thres}
{\rm minimize}~~~\|X\|_0\\
{\rm subject~to}~~~Y=AX+N, \notag
\end{eqnarray}
where $A\in\mathbb{R}^{m\times n}$ is a random matrix with i.i.d.
$\mathcal{N}(0,1/m)$ entries, $N
=[\nb_1,\cdots,\nb_r]\in\mathbb{R}^{m\times r}$ is  an additive
noise matrix whose components have i.i.d. $\mathcal{N}(0,1/m)$
entries, and $m=m(n)\nearrow\infty, k=k(n)\nearrow\infty$ and
$r=r(n)\nearrow\infty$ such that  $k/m<1-\epsilon$, $r/k<1-\epsilon$
for some $\epsilon>0$ and ${\rm rank}(AX)=r\leq k=\|X\|_0$. Here, we
assume that $\rho:=\lim_{n\rightarrow\infty}m(n)/n>0$ and
$\alpha=\lim_{n\rightarrow\infty}r(n)/k(n)\geq 0$ exist.
\end{definition}

\bigskip

Note that the conditions $k/m<1-\epsilon$, and
$r/k<1-\epsilon$ are technical conditions that prevent $m,k$, and
$r$ from reaching equivalent values when $n\rightarrow \infty$.

\subsection{Sufficient condition for generalized MUSIC}
For the case of a noisy measurement, $Y$ is corrupted and the
corresponding noise subspace estimate $Q$ is not correct. However,
the following theorem shows that if the $I_{k-r} \subset {\rm
supp}X$, then the generalized MUSIC estimate is consistent and
achieves the correct estimation of the remaining $r$-indices for
sufficiently large SNR.

\begin{theorem}\label{genmusic-noisy}
For a $\mathrm{LSMMV}(m,n,k,r;\epsilon)$, if we have $I_{k-r}\subset {\rm supp}X$, then we can find remaining $r$ indices of supp$X$ with the generalized MUSIC criterion if
\begin{equation}\label{num-genmusic}
m\geq \max\left\{k(1+\delta)\left[1-\frac{4(\kappa(B)+1)}{{\sf SNR}_{\min}(Y)-1}\right]^{-1},(1+\delta)(2k-r+1)\right\}
\end{equation}
for some $\delta>0$ provided that ${\sf
SNR}_{\min}(Y):=\sigma_{\min}(B)/\|N\|> 1+{4(\kappa(B)+1)}$, where
$\kappa(B)$ denotes the condition number and $\sigma_{\min}(B)$
denotes the smallest singular value of $B$.
\end{theorem}
\begin{proof}
See Appendix D.
\end{proof}

Note that for ${\sf SNR}_{\min}(Y)\rightarrow\infty$, the condition becomes
$m\geq (1+\delta)(2k-r+1)$ for some $\delta>0$. However, as ${\sf SNR}_{\min}(Y)$ decreases, the first term dominates and we need more sensor elements.

\subsection{Sufficient condition for partial support recovery using 2-thresholding} Now, define the thresholding estimate as $I_t=\{p_i\}_{i=1}^{k-r}$
where
$$\rho(j)=\|\mathbf{a}_j^*Y\|_F^2.$$ Now, we derive sufficient conditions for the success of 2-thresholding in detecting $k-r$ support when $r$ is a small fixed number or when $r$ is proportionally increasing with respect to $k$.
\begin{theorem}\label{lem:num-thres}
For a $\mathrm{LSMMV}(m,n,k,r;\epsilon)$, suppose ${\sf MSR}_{min}^{(k-r)}$ are deterministic sequences and
\begin{equation}\label{snr-thres}
{\sf SNR}_{\min}(Y)>
\frac{2\kappa(B)+\sqrt{4\kappa(B)^2+2r{\sf MSR}_{\min}^{(k-r)}/(\sigma_{\min}^2(B))}}{r{\sf MSR}_{\min}^{(k-r)}/(\sigma_{\min}^2(B))},
\end{equation}
\begin{equation}\label{num-thres1}
m>2(1+\delta)
\frac{\left(\frac{\|X\|_F}{\sqrt{r}}\sqrt{\log{(k-r)}}+\sqrt{\frac{B(n,k,r)}{r}}\right)^2}{\left(\sqrt{\sf{MSR}_{\min}^{(k-r)}}-\sqrt{(2(2\|B\|+\|N\|)\|N\|)/r}\right)^2}
\end{equation}
where
\begin{eqnarray}\label{bnkr-def}
B(n,k,r)=
\left\{\begin{array}{ll}
\sigma_{\min}^2(B)\log{(n-k)}+(\|B\|_F^2-r\sigma_{\min}^2(B))\log{((n-k)r)},&\\
\quad\quad\quad\quad\quad\quad\quad\quad\quad\quad\quad\quad\quad\quad{\rm if}~r{~\rm is~a~fixed~positive~integer}&\\
&\\
\sigma_{\min}^2(B)\frac{r}{2}+(\|B\|_F^2-r\sigma_{\min}^2(B))\log{((n-k)r)},&\\
\quad\quad\quad\quad\quad\quad\quad\quad\quad\quad\quad\quad\quad\quad{\rm if}~\alpha:=\lim\limits_{n\rightarrow\infty}r/k>0&
\end{array}
\right.
\end{eqnarray}
where
$${\sf MSR}_{\min}^{k-r}=\frac{\|X\|_{(k-r)}^2}{r},$$
and $\|X\|_{(k-r)}^2$ is the ($k-r$)-th value if we are ordering the values of $\|\xb^i\|^2$ for $1\leq i\leq n$ with descending order.
Then, 2-thresholding
asymptotically finds a $k-r$ sparsity pattern.
\end{theorem}
\begin{proof}
See Appendix F.
\end{proof}

\begin{itemize}
\item  For noiseless single measurement vector (SMV) case, i.e. $r=1$, if ${\sf SNR}_{\min}(Y)\rightarrow\infty$, this becomes
\begin{equation*}
m>2(1+\delta)\frac{\left(\|\xb\|\sqrt{\log{(k-1)}}+\|\bb\|\sqrt{\log{(n-k)}}\right)^2}{\min\limits_{j\in I_t}|x_j|^2}.
\end{equation*}
Using Lemma \ref{lem:xb-equiv} in Appendix E, we have
\begin{equation}\label{xb-equiv}
\lim\limits_{n\rightarrow\infty}\frac{\|\bb\|^2}{\|\xb\|^2}=1.
\end{equation}
Hence, we have
\begin{equation*}
m\geq 2(1+\delta)\frac{\|\xb\|^2}{\min\limits_{j\in I_t}|x_j|^2}\left(\sqrt{\log{(k-1)}}+\sqrt{\log{(n-k)}}\right)^2
\end{equation*}
for some $\delta>0$, as the sufficient condition for 2-thresholding in SMV cases.
Compared to the result in \cite{fletcher2009necessary} as
\begin{equation*}
m\geq 2(1+\delta)\frac{\|\xb\|^2}{\min\limits_{j\in {\rm supp}X}|x_j|^2}(\sqrt{\log{k}}+\sqrt{\log{(n-k)}})^2,
\end{equation*}
our bound has a slight gain due to $\sqrt{\log{(k-1)}}$ and  $\min_{j\in I_t}|x_j|^2$, where $|I_t|=k-1$. This is because even for  the SMV problem, the one remaining  index can be estimated using the generalized MUSIC criterion.
\item If $\|B\|_F^2=r\sigma_{\min}^2(B)$, $r$ is a fixed number and ${\sf SNR}_{\min}(Y)\rightarrow\infty$, then our bound can be reduced as
\begin{eqnarray*}
m&\geq& 2(1+\delta)\frac{\left(
\frac{\|X\|_F}{\sqrt{r}}\sqrt{\log{(k-r)}}+\frac{\sqrt{r\sigma_{\min}^2(B)}}{r}\sqrt{\log{(n-k)}}\right)^2}{{\sf MSR}_{\min}^{k-r}},
\end{eqnarray*}
when the measurement is noiseless.
Using Lemma \ref{lem:xb-equiv} in Appendix E, this can be simplified as
\begin{equation}\label{num-thres-case2}
m\geq 2(1+\delta)
\frac{\|X\|_F^2}{\|X\|_{(k-r)}^2}\left(\sqrt{\log{(k-r)}}+\sqrt{\frac{\log{(n-k)}}{r}}\right)^2.
\end{equation}
Therefore, the MMV gain over SMV mainly comes from $\sqrt{(\log{(n-k)})/r}$.
\item If $\|B\|_F^2=r\sigma_{\min}^2(B)$ and $\lim_{n\rightarrow\infty}r/k=\alpha>0$, then under the condition \eqref{xb-equiv} we have
\begin{equation*}
m\geq 2(1+\delta)
\frac{\|X\|_F^2}{\|X\|_{(k-r)}^2}\left(\sqrt{\log{(k-r)}}+\frac{1}{\sqrt{2}}\right)^2.
\end{equation*}
Therefore, the $\log{(n-k)}$ factor disappears, which provides more MMV gain compared to \eqref{num-thres-case2}.
\end{itemize}

\subsection{Sufficient condition for partial support recovery using subspace S-OMP} Next, we consider the minimum number of measurements for compressive MUSIC with S-OMP. In analyzing S-OMP, rather than analyzing the distribution of  $\|\ab_j^{*}P_{R(A_{I_t})}^{\perp}B\|_F^2$ where $I_t$ denotes the set of indices which are chosen in the first $t$ step of S-OMP,  we consider the following version of subspace S-OMP  due to its superior performance \cite{DaviesEldar2010,LeeBresler2010}.

\begin{enumerate}
\item Initialize $t=0$ and $I_0=\emptyset$.
\item Compute $P^\perp_{R(A_{I_t})}$ which is the projection operator onto the orthogonal complement of the span of $\{\ab_j:j\in I_t\}$.
\item Compute $P^\perp_{R(A_{I_t})}B$ and for all $j=1,\cdots, n$, compute
$\rho(t,j)=\|\ab_j^{*}P_{{R(P^\perp_{R(A_{I_t})}B)}}\|_F^2$.
\item Take $j_t=\arg\max_{j=1,\cdots,n}\rho(t,j)$ and $I_{t+1}=I_t\cup
\{j_t\}$. If $t<k$ return to Step 2.
\item The final estimate of the sparsity pattern is $I_k$.
\end{enumerate}

Now, we also consider two cases according to the number of multiple measurement vectors. First, we consider the case when the number of multiple measurement vectors is a finite fixed number. Conventional compressive sensing (the SMV problem) is  this kind of  case. Second, we consider the case when $r$ is proportional to $n$. This case includes the conventional MUSIC case.

\begin{theorem}\label{lem:num-somp1}
For $\mathrm{LSMMV}(m,n,k,r;\epsilon)$, let ${\sf SNR}_{\min}(Y)=\sigma_{\min}(AX)/\|N\|$ and
suppose the following conditions
hold: \\
(a) $r$ is a fixed finite number.\\
(b) Let ${\sf SNR}_{\min}(Y)$ satisfy
\begin{equation}\label{snr:somp:rsmall}
{\sf SNR}_{\min}(Y)>1+\frac{4k}{r}(\kappa(B)+1).
\end{equation}
If we have
\begin{equation}\label{num:somp:rsmall}
m>k(1+\delta)\left[1-\frac{4k}{r}
\frac{(\kappa(B)+1)}{{\sf SNR}_{\min}(Y)-1}\right]^{-1}\frac{2\log{(n-k)}}{r},
\end{equation}
then we can find $k-r$ correct indices of ${\rm supp}X$ by applying subspace S-OMP.
\end{theorem}
\begin{proof}
See Appendix G.
\end{proof}
\begin{itemize}
\item As a simple corollary of Theorem~\ref{lem:num-somp1}, when ${\sf SNR}_{\min}(Y)\rightarrow\infty$, we can
easily show that the number of sensor elements required for the
conventional OMP to find the all $k$-support indices in SMV problem is given by
\begin{equation}\label{eq-num-2thres}
m>2(1+\delta)k\log{(n-k)}\ ,
\end{equation}
for a small $\delta>0$. This is equivalent to the result in
\cite{fletcher2009necessary}.

\item When ${\sf SNR}_{\min}(Y)\rightarrow\infty$, then the number of sensor elements for subspace S-OMP is
$$m>2(1+\delta)\frac{k}{r}\log{(n-k)}$$ for some $\delta>0$. Hence, the sampling ratio is the reciprocal of the number of multiple measurement vectors.

\item Since $k\rightarrow\infty$ in our large system model,  \eqref{snr:somp:rsmall} tells us that the required ${\sf SNR}_{\min}(Y)$ should increase to infinity.

\end{itemize}

Next, we consider the case that $r$ is proportionally increasing with respect $k$. In this case, we have the following theorem.

\begin{theorem}\label{lem:num-somp2}
For $\mathrm{LSMMV}(m,n,k,r;\epsilon)$, let ${\sf SNR}_{\min}(Y)=\sigma_{\min}(AX)/\|N\|$ and
suppose the following conditions
hold. \\
(a) $r$ is proportionally increasing with respect to $k$ so that $\alpha:=\lim_{n\rightarrow\infty}r(n)/k(n)>0$ exist.\\
(b) Let ${\sf SNR}_{\min}(Y)$ satisfy
\begin{equation}\label{snr:somp:rlarge}
{\sf SNR}_{\min}(Y)>1+\frac{4}{\alpha}(\kappa(B)+1).
\end{equation}
Then if we have
\begin{equation}\label{num:somp:rlarge}
m>k(1+\delta)^2\frac{1}{\left[1-\frac{4}{\alpha}\frac{\kappa(B)+1}{{\sf SNR}_{\min}(Y)-1}\right]^2}\left[2-F(\alpha)\right]^2,
\end{equation}
for some $\delta>0$ where
$$F(\alpha)=\frac{1}{\alpha}\int_0^{4t_1(\alpha)^2}xd\lambda_1(x),$$
$d\lambda_1(x)=(\sqrt{(4-x)x})/(2\pi x)$
is the probability measure with support $[0,4]$, $0\leq t_1(\alpha)\leq 1$ satisfies
$\int_0^{2t_1(\alpha)}ds_1(x)=\alpha$
and
$ds_1(x)=(1/\pi)\sqrt{4-x^2}$ is a probability measure with support $[0,2]$.
Here, $F(\alpha)$ is an increasing function such that $F(1)=1$ and $\lim_{\alpha\rightarrow0^{+}}F(\alpha)=0$. Then we can find $k-r$ correct indices of ${\rm supp}X$ by applying subspace S-OMP.
\end{theorem}
\begin{proof}
See Appendix G.
\end{proof}
\begin{itemize}
\item As a corollary of Theorem \ref{lem:num-somp2}, when $r(n)/k(n)\rightarrow 1$ and ${\sf SNR}_{\min}(Y)\rightarrow\infty$, we can see that the number of sensor elements required for subspace S-OMP to find $k-r$ support indices is given by
\begin{equation*}
m > (1+\delta)k,
\end{equation*}
for a small $\delta>0$, which is the same as the number of sensor elements required for MUSIC.
\item We can expect that the number of sensor elements required for subspace S-OMP to find $k-r$ support indices is at most $4(1+\delta)k$ in the noiseless case, where $\delta>0$ is an arbitrary small number. Hence, the $\log{n}$ factor is not necessary.
\item Unlike the case in Theorem \ref{lem:num-somp1}, the {\sf SNR} condition is now lower bounded by a finite number $1+(4/\alpha)(\kappa(B)+1)$. This implies that we don't need infinite {\sf SNR} for support recovery, in contrast to SMV or Theorem \ref{lem:num-somp1}. This is one of the important advantages of MMV over SMV.
\end{itemize}

\section{Numerical Results}\label{sec:simulation}
\setcounter{equation}{0} \setcounter{theorem}{0} \indent

In this section, we demonstrate the performance of compressive MUSIC.
This new algorithm is compared to the conventional MMV algorithms,
especially 2-SOMP, 2-thresholding and $l_{2,1}$ mixed-norm approach \cite{malioutov2005ssr}. We do not compare the new algorithm with the
classical MUSIC algorithm since it fails when $r<k$. We declared the
algorithm as a success if the estimated support is the same as the true
${\rm supp}X$, and the success rates were averaged for $5000$
experiments. The simulation parameters were as follows: $m\in\{1, 2,
\ldots, 60\},~ n=200,~ k\in\{1, 2, \ldots, 30\}$, and $r\in\{1, 3, 8,
16\}$, respectively. Elements of sensing matrix $A$ were generated
from a Gaussian distribution having zero mean and variance of $1/m$, and
the ${\rm supp}X$ were chosen randomly. The maximum iteration was
set to $k$ for the S-OMP algorithm.

\begin{figure}[htbp]
 \centerline{\epsfig{figure=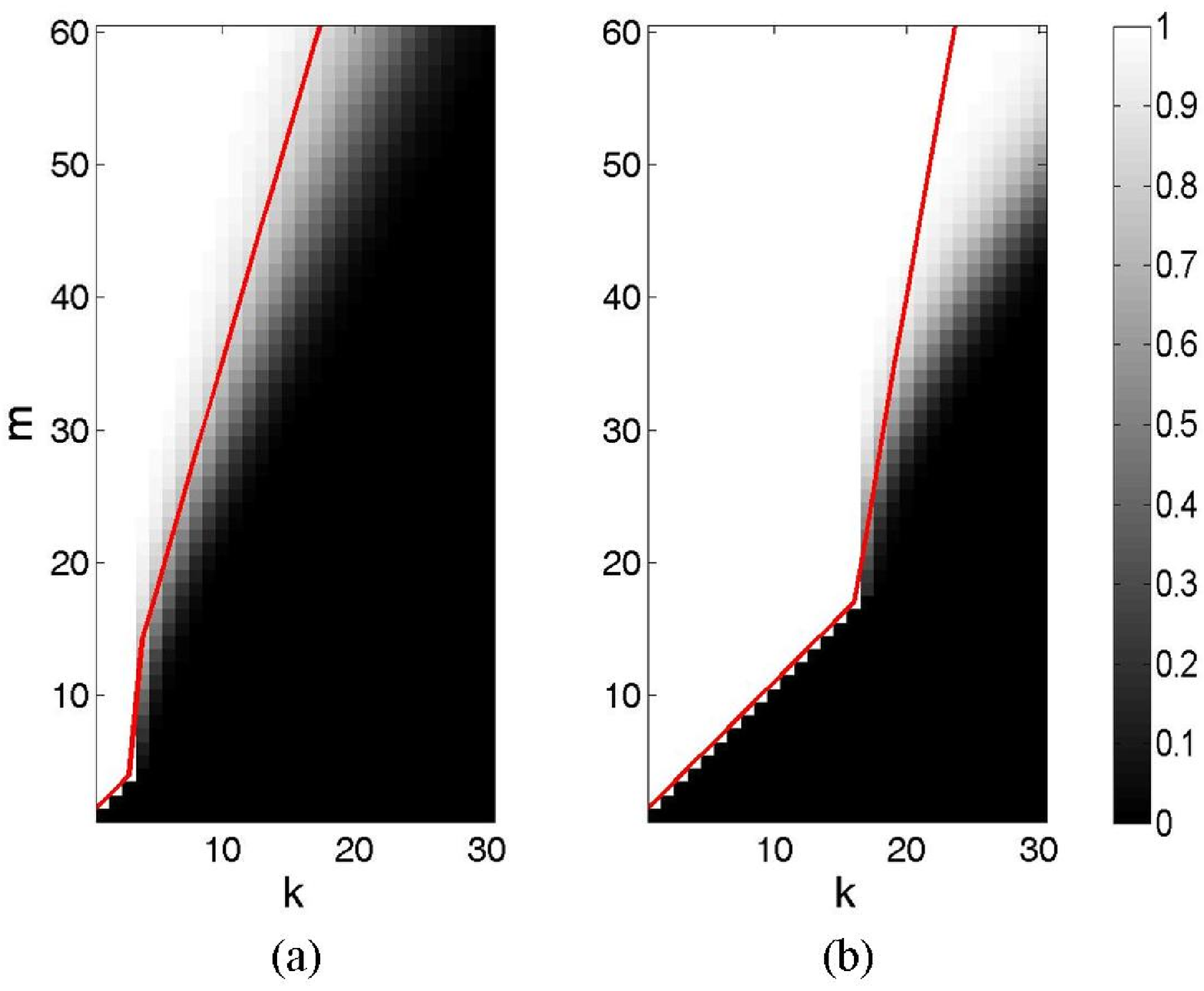,width=8cm}}
  \caption{ Phase transition map for compressive MUSIC with subspace S-OMP when $n=200$, $\textsf{SNR}=\infty$, $\|\xb^i\|^2$ is constant for all $i=1,\cdots,n$, and
  (a) $r=3$, and (b) $r=16$.
  The overlayed curves are calculated based on \eqref{eq:phase_somp}. }
 \label{fig:bound_CMUSIC_SOMP}
\end{figure}

 According
to \eqref{num-genmusic}, \eqref{num:somp:rsmall} and \eqref{num:somp:rlarge}, for noiseless measurements, piece-wise continuous boundaries exist for the phase
transition of CS-MUSIC with subspace S-OMP:
\begin{equation}\label{eq:phase_somp}
   m > \left\{
          \begin{array}{ll}
           k+1 , & \hbox{$r=k$;} \\
           2k-r+1 , & \hbox{$r<k$;} \\
           (2k\log{(n-k)})/r & \hbox{$r\ll k$.}\\
           k[2-F(\alpha)]^2, & \hbox{$\lim_{n\rightarrow \infty} r/k> 0$.}
          \end{array}
        \right.
\end{equation}
Note that in our canonical MMV model, $r=k$ includes many MMV
problems in which the number of snapshots is larger than the sparsity level
since our canonical MMV model reduces the effective snapshot $r$ as
$r\geq k$.
 Figure~\ref{fig:bound_CMUSIC_SOMP}(a) shows a typical phase transition map of our
compressive MUSIC with subspace S-OMP for noiseless measurements when $n=200$
and $r=3$ and $\|\xb^i\|$ is constant for all $i=1,\cdots,n$.
Even though the simulation step is not in the large system regime, but $r$ is quite small, so that we can expect that $(2k\log{(n-k)})/r$ is a boundary for phase transition.
Figure~\ref{fig:bound_CMUSIC_SOMP}(b) corresponds to the case when
$r=16$ and $\|\xb^i\|$ is constant for all $i=1,\cdots,n$.  Since  in this setup $r$ is comparable to $k$,  we use the $k[2-F(\alpha)]^2$ as a boundary.
The results  clearly indicates the
tightness of our sufficient condition.

\begin{figure}[htbp]
 \centerline{\epsfig{figure=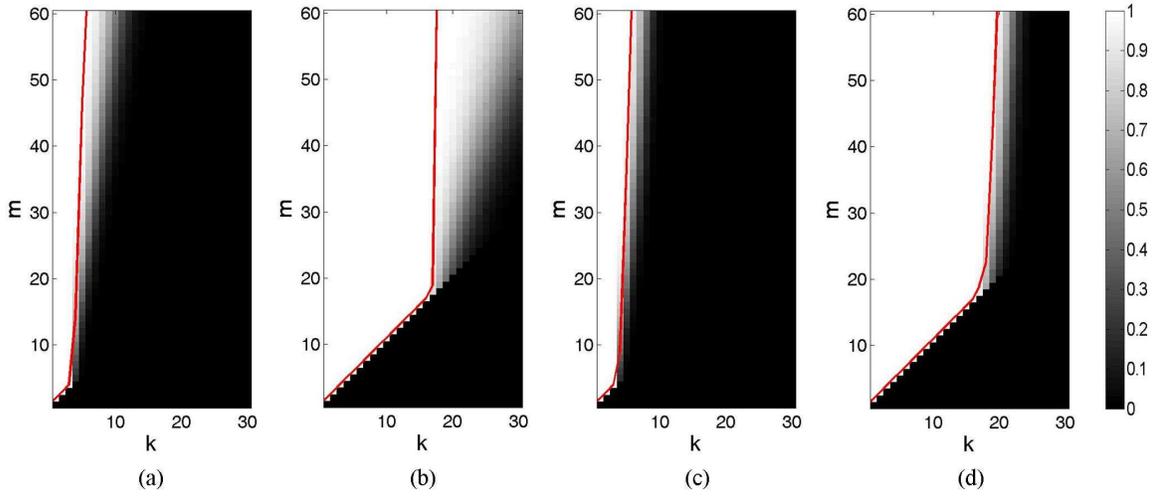,width=16cm}}
  \caption{ Phase transition map for compressive MUSIC with 2-thresholding when $n=200$, $\textsf{SNR}=\infty$, and
  (a) $r=3$, (b) $r=16$ when $\|\xb^i\|^2$ is constant for all $i=1,\cdots,n$, and (c) $r=3$, (d) $r=16$ when $\|\xb^i\|^2=0.7^{i-1}$.
  The overlayed curves are calculated based on \eqref{eq:phase_thresholding}. }
 \label{fig:bound_CMUSIC_Pth}
\end{figure}

Similarly,  multiple piecewise continuous
boundaries exist for the phase transition map for compressive MUSIC with
2-thresholding:
\begin{equation}\label{eq:phase_thresholding}
    m > \left\{
          \begin{array}{ll}
           k+1 , & \hbox{$r= k$;} \\
           2k-r+1 , & \hbox{$r<k$;} \\
           2\left(\|X\|_F\sqrt{\log{(n-k)}}/\sqrt{r}+\sqrt{B(n,k,r)/r}\right)^2/{\sf MSR}_{\min}^{k-r},& \hbox{$r<k$.}
          \end{array}
        \right.
\end{equation}
Since the phase transition boundary depends on the unknown joint sparse signal $X$ through $\|X\|_F$ and ${\sf MSR}_{\min}^{k-r}$, we investigate this effect.
 Figure~\ref{fig:bound_CMUSIC_Pth}(a) and (b) show a typical phase transition map of our
compressive MUSIC with 2-thresholding when $r=3$ and $r=16$, respectively, for noiseless measurements and $\|\xb^i\|$ are constant for all $i$;
Figure~\ref{fig:bound_CMUSIC_Pth}(c) and (d) corresponds to the same case except $\|\xb^i\|^2=(0.7)^i$.
We overlayed theoretically calculated
phase boundaries over the phase transition diagram.   The empirical phase transition diagram clearly revealed the effect of the distribution $X$.
Still, the theoretically calculated boundary clearly indicates the
tightness of our sufficient condition.

\begin{figure}[htbp]
 \centerline{\epsfig{figure=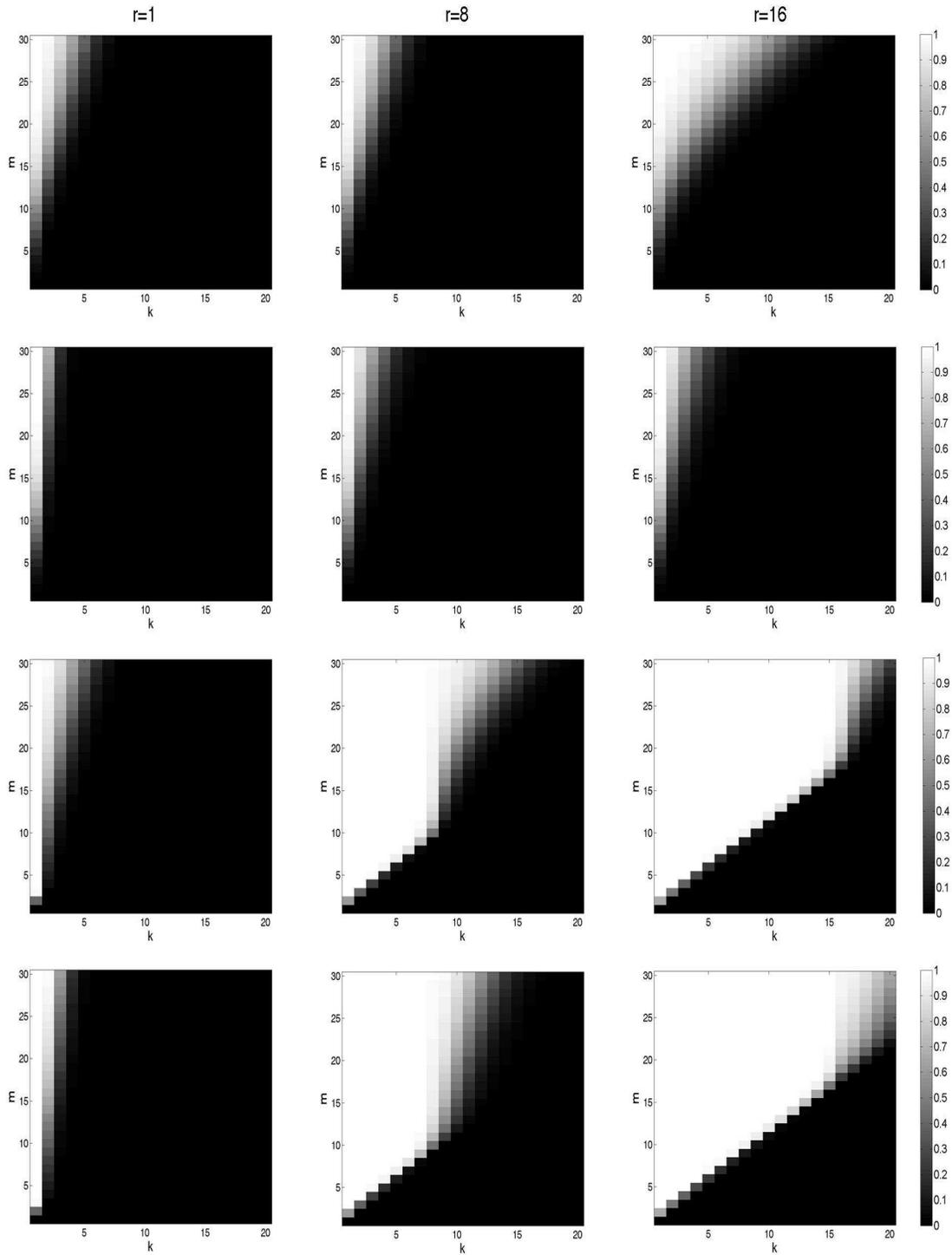,width=15cm, height=20cm}}
  \caption{Recovery rates for various $m$ and $k$ when $\textsf{SNR}=$40dB and non-zero rows of
  $\|\xb^i\|$ are constant for all $i$.
  Each row (from top to bottom) indicates the recovery rates by S-OMP, 2-thresholding, and compressive MUSIC with subspace S-OMP and 2-thresholding.
  Each column (from left to right) indicates $r=1,8$ and $16$, respectively. }
 \label{fig:all_SNR40_uni}
\end{figure}

Fig. ~\ref{fig:all_SNR40_uni} shows the success rate of S-OMP,
2-thresholding, and compressive MUSIC with subspace S-OMP and 2-thresholding
for 40dB noisy measurement when $\|\xb^i\|$ is constant for all
$i=1,\cdots,n$. When $r=1$, the performance level of the compressive
MUSIC algorithm is basically the same as that of a compressive sensing
algorithm such as 2-thresholding and S-OMP. When $r=8$, the recovery
rate of the compressive MUSIC algorithm is higher than the case $r=1$,
and the compressive MUSIC algorithm outperforms the conventional
compressive sensing algorithms. If we increase $r$ to 16, the
success of the compressive MUSIC algorithm becomes nearly deterministic
and approaches the $l_0$ bound, whereas conventional compressive
sensing algorithms do not.


%

\begin{figure}[htbp]
 \centerline{\epsfig{figure=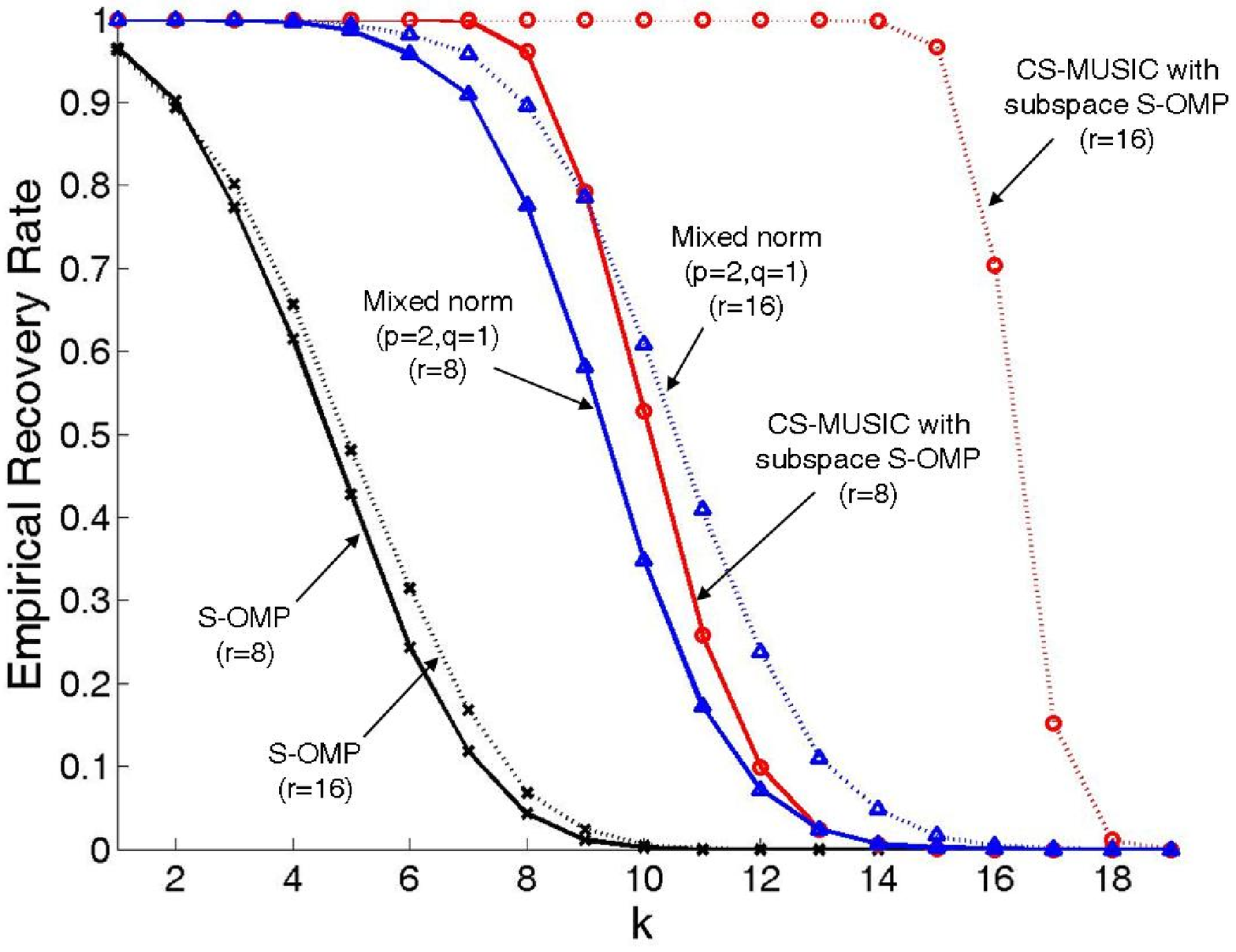,width=8cm}\epsfig{figure=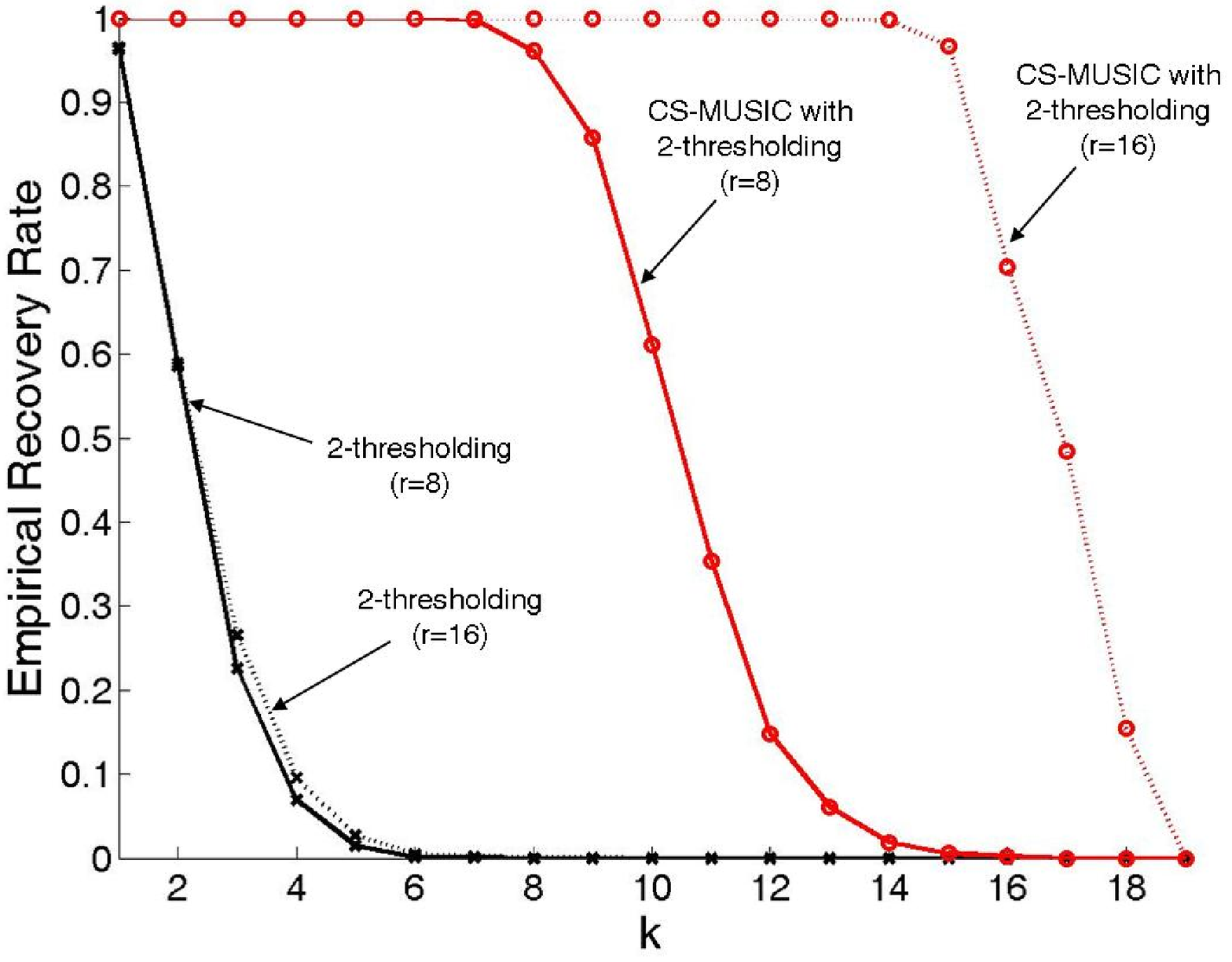,width=8cm}}
 \centerline{ \mbox{(a)}\hspace{8cm}\mbox{(b)}  }
  \caption{Recovery rates by various MMV algorithms for a uniform source when $n=200,~m=20,~ r=8$, and $16$ and $\textsf{SNR}=$40dB:
  (a) recovery rate for S-OMP, compressive MUSIC with subspace S-OMP, and mixed norm approach when $p=2,q=1$ and (b) recovery rate for 2-thresholding and compressive MUSIC with 2-thresholding.
  }
 \label{fig:m20_uni}
\end{figure}

In order to compare compressive MUSIC with other methods more
clearly the recovery rates of various algorithms are plotted in
Fig. \ref{fig:m20_uni}(a) for S-OMP, compressive MUSIC with subspace S-OMP, and the mixed norm approach when $p=2,q=1$; and in Fig. ~\ref{fig:m20_uni}(b) for 2-thresholding and compressive MUSIC with 2-thresholding,
when $n=200$, $m=20$ and $r=8,16$, $\|\xb^i\|$ is constant, and $\textsf{SNR}=40$dB. Note that
compressive MUSIC outperforms the existing methods.

\begin{figure}[htbp]
 \centerline{\epsfig{figure=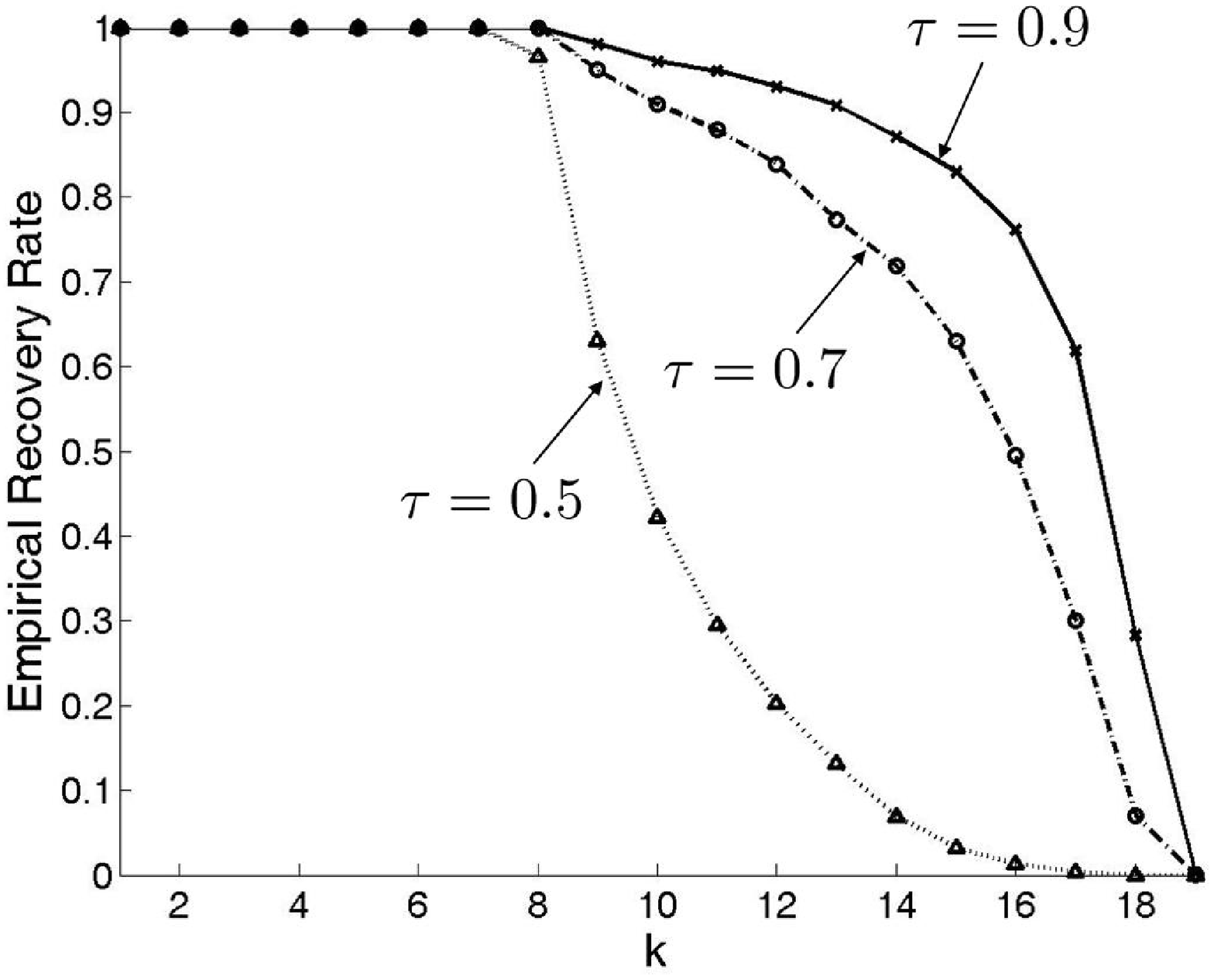,width=5.4cm}\epsfig{figure=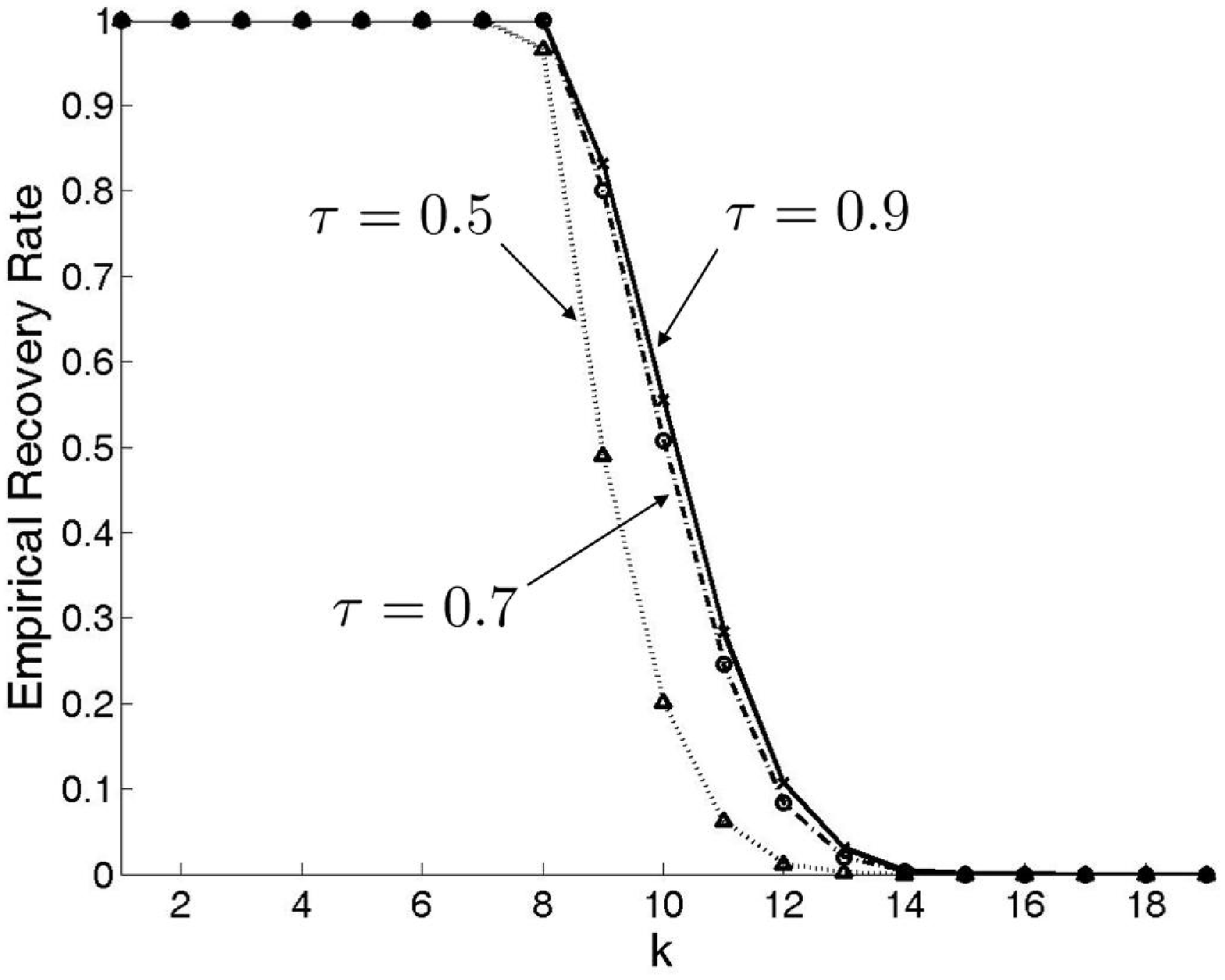,width=5.4cm}\epsfig{figure=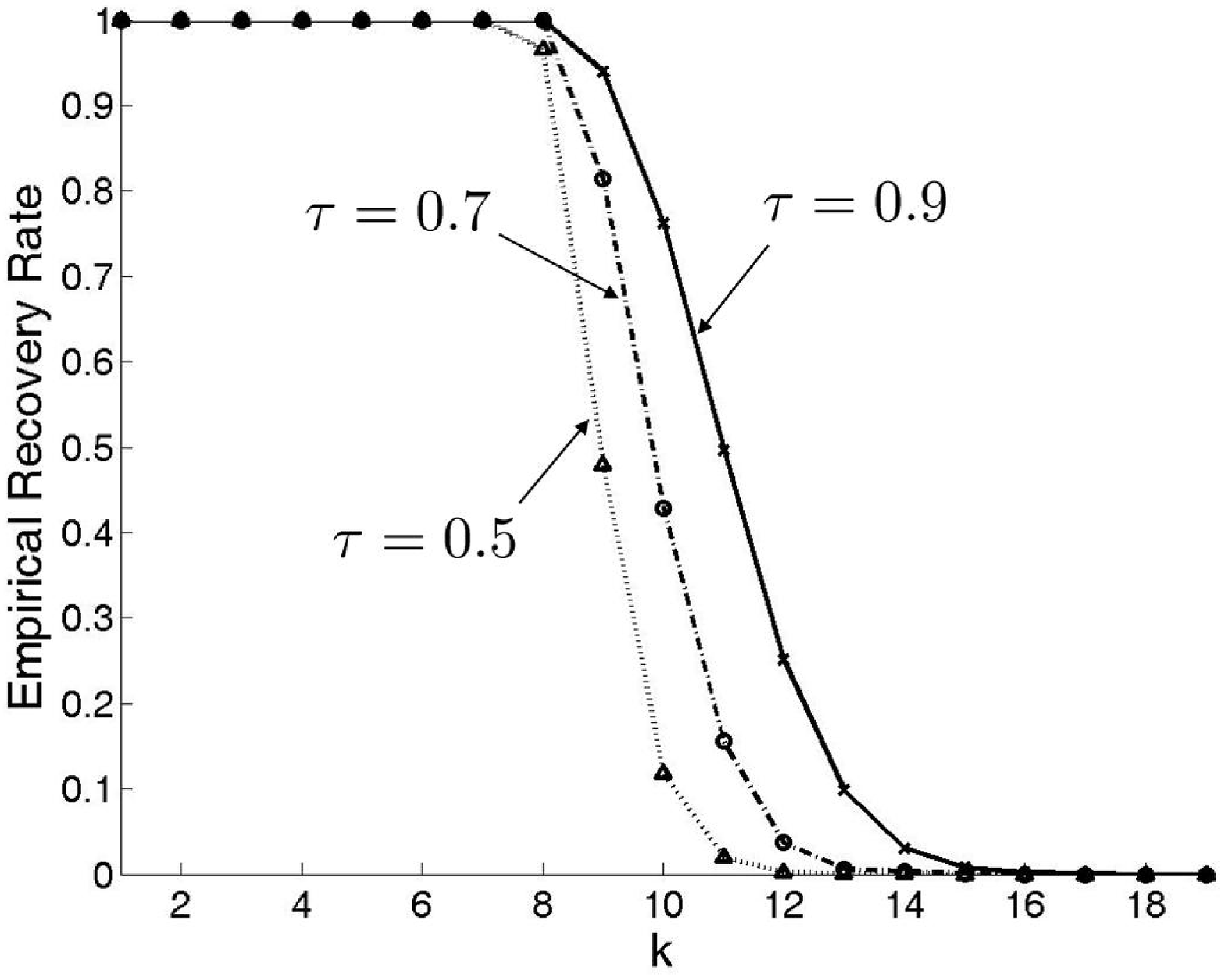,width=5.4cm}}
 \centerline{ \mbox{(a)}\hspace{5.4cm}\mbox{(b)}\hspace{5.4cm}\mbox{(c)}  }
  \caption{ Recovery rates by compressive MUSIC when $k-r$ nonzero supports are estimated by (a) an ``oracle" algorithm, (b) subspace S-OMP, and (c) 2-thresholding. Here, $X$ is given with $\tau=0.9$, $\tau=0.7$ and $\tau=0.5$. Smaller $\tau$ provides larger condition number $\kappa(X)$. The measurements are corrupted by additive Gaussian noise of $\textsf{SNR}=$40dB and $n=200,~m=20$, $r=8$.}
 \label{fig:40db-cond}
\end{figure}

To show the relationship between the recovery performance in the noisy setting and the condition number of matrices $X$, we performed the simulation on the recovery results for three different types of the source model $X$. More specifically, the singular values of $X$ are set to be exponentially decaying with (i) $\tau=0.9$, (ii) $\tau=0.7$ and (iii) $\tau=0.5$ respectively, i.e. the singular values of $X$ are given by $\sigma_j=\tau^{j-1}$ for $j=1, \cdots, {\rm rank}(X)$. In this simulation, we are using noisy samples that are corrupted by additive Gaussian noise of $\textsf{SNR}=$40dB. Figure \ref{fig:40db-cond}(a) shows the results when $k-r$ entries of the support are known {\em a priori} by an ``oracle" algorithm, whereas $k-r$ entries of the support are determined by subspace S-OMP  in Fig. \ref{fig:40db-cond}(b) and by thresholding in Fig. \ref{fig:40db-cond}(c). The results provide evidence of the significant impact of the condition number of $X$.

\begin{figure}[htbp]
 \centerline{\epsfig{figure=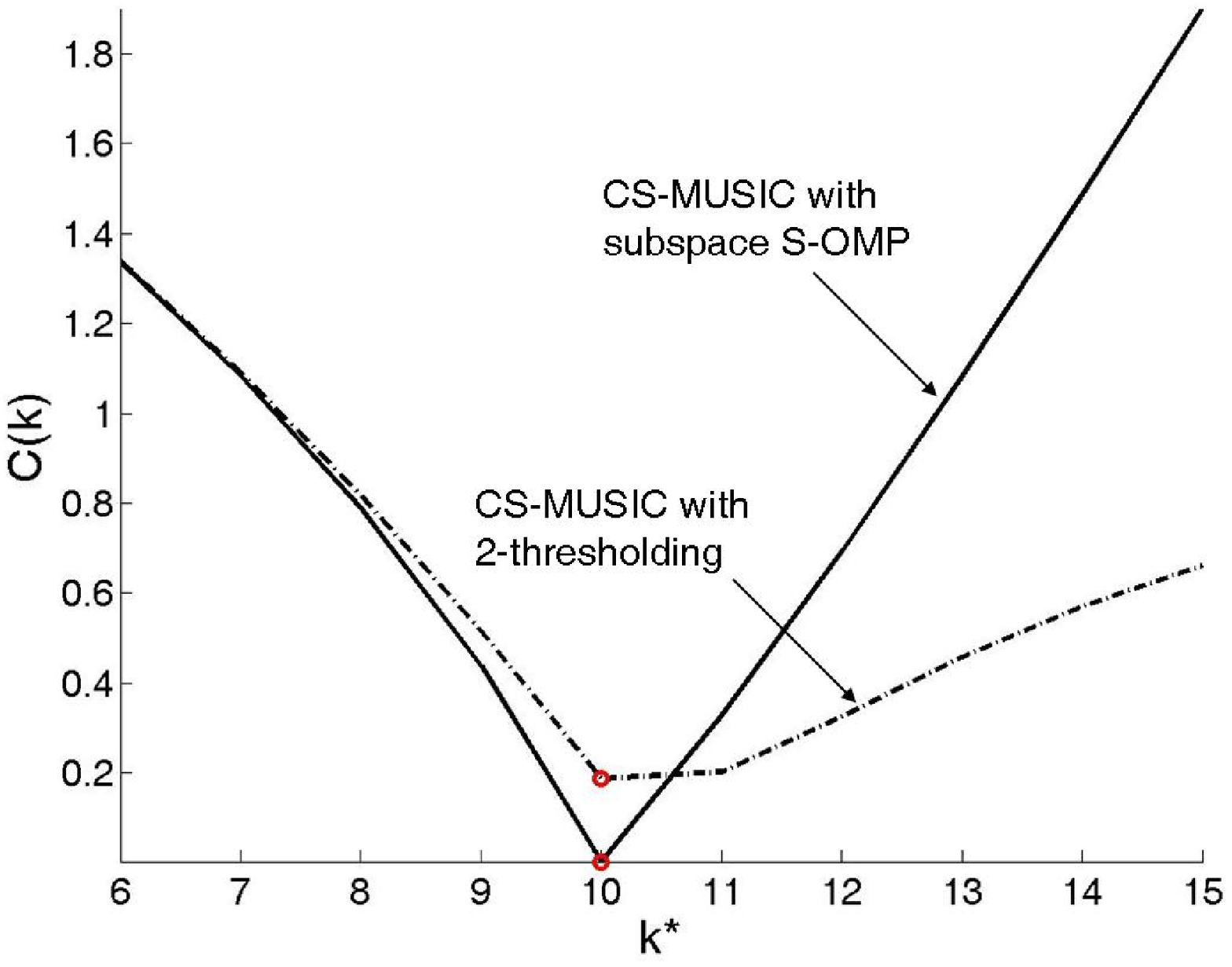,width=8cm}\epsfig{figure=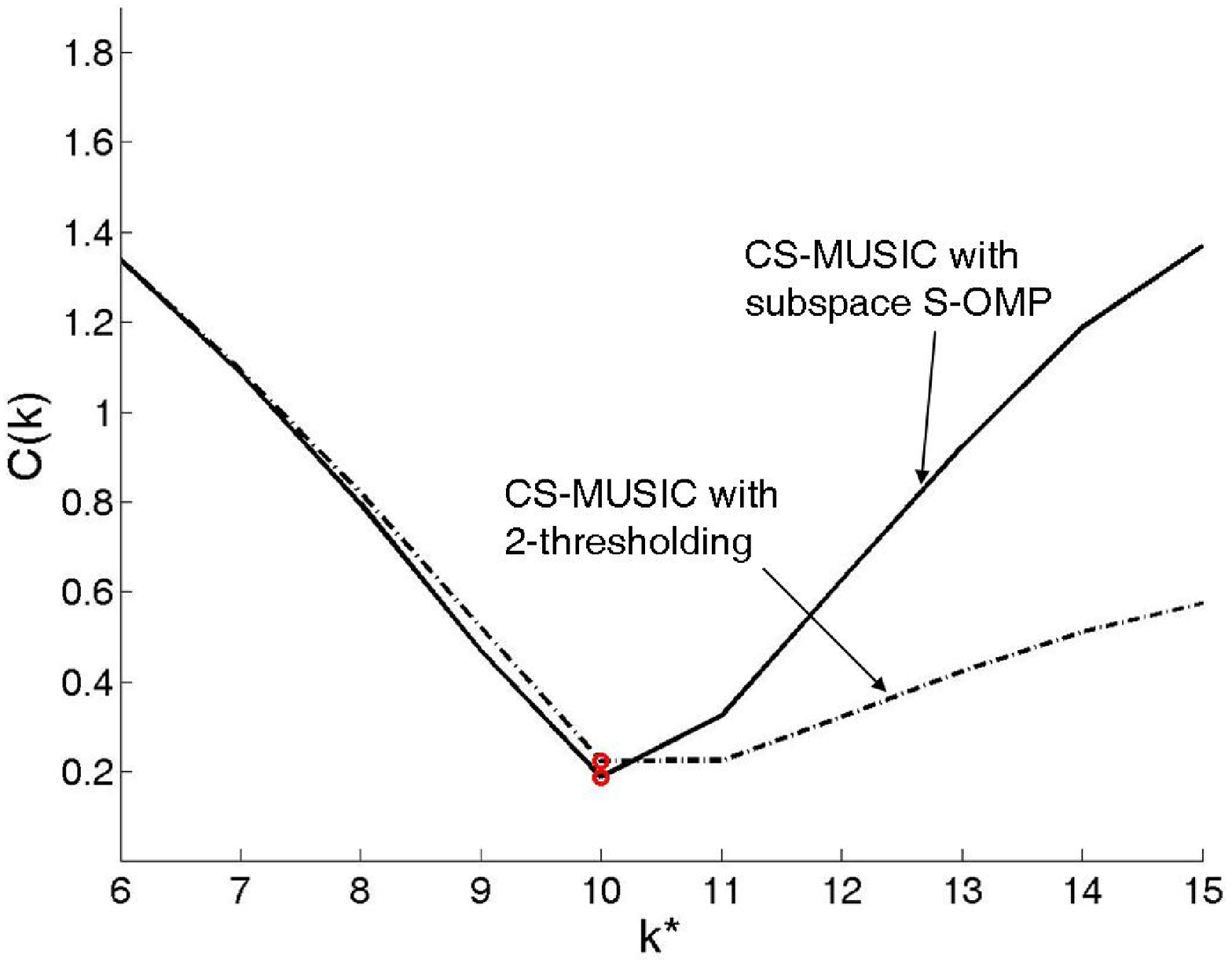,width=8cm}}
 \centerline{ \mbox{(a)}\hspace{8cm}\mbox{(b)}  }
  \caption{ Cost function for sparsity estimation when $n=200,~m=40,~r=5,~k=10$,
  and the measurements are
  (a) noiseless and (b) corrupted by additive Gaussian noise of $\textsf{SNR}=$40dB. The circles illustrate the local minima,
  whose position corresponds to the true sparsity level.}
 \label{fig:estimation_k}
\end{figure}

Figure \ref{fig:estimation_k} illustrates the cost function to
estimate the unknown sparsity level, which confirms that compressive
MUSIC can accurately estimate the unknown sparsity level $k$ as
described in Lemma \ref{lem-supp-est}. In this simulation, $n=200,~m=40$ and $r=5$. The correct support size $k$ is marked as circle. Note that
$C({\hat k})$ has the smallest value at that point for the noiseless
measurement cases, as shown Fig. ~\ref{fig:estimation_k}(a),
confirming our theory. For the 40dB noisy measurement case, we can still
easily find the correct $k$ since it corresponds to the first local
minimizer as ${\hat k}$ increases, as shown in Fig.~\ref{fig:estimation_k}(b).

\section{Discussion}\label{sec:dis}

\subsection{Comparison with subspace-augmented MUSIC \cite{LeeBresler2010}}
Recently, Lee and Bresler \cite{LeeBresler2010} independently
developed a hybrid MMV algorithm called as subspace-augmented MUSIC
(SA-MUSIC). The SA-MUSIC performs the following procedure.
 \begin{enumerate}
   \item Find $k-r$ indices of ${\rm supp}X$ by applying SOMP to the MMV problem $U=AX$ where the set of columns of $U$ is an orthonormal basis for $R(B)$.
   \item Let $I_{k-r}$ be the set of indices which are taken in Step 1 and $S=I_{k-r}$.
   \item For $j\in \{1,\cdots,n\}\setminus I_{k-r}$, compute $\eta(j)=\|\tilde{Q}^{*}\ab_j\|^2$ where $\tilde{Q}\in \mathbb{R}^{m \times (m-k)}$ consists of orthonormal columns such that $\tilde{Q}^{*}[U~~A_{I_{k-r}}]=0$.
   \item Make an ascending ordering of $\eta(j)$, $j\notin I_{k-r}$, choose indices that correspond to the first $r$ elements, and put these indices into $S$.
\end{enumerate}
By Theorem \ref{cmusic-geo}(c), we can see that the
subspace-augmented MUSIC is equivalent to compressive MUSIC, sinces
the MUSIC criterion in subspace-augmented measurement
$[U~A_{I_{k-r}}]$ and the generalized MUSIC criterion in compressive
MUSIC are equivalent. Therefore, we can expect that the performance
of both algorithm should be similar except for the following
differences. First, the subspace S-OMP in Lee and Bresler
\cite{LeeBresler2010} is applying the subspace decomposition once
for the data matrix $Y$ whereas our analysis for the subspace S-OMP
is based on subspace decomposition for the residual matrix at each
step. Hence, our subspace S-OMP is more similar to that of
\cite{DaviesEldar2010}. However, based on our experiments the two
versions of the subspace S-OMP provide similar performance when
combined with the generalized MUSIC criterion. Second, the
theoretical analysis of SA-MUSIC is  based on the RIP condition
whereas ours is based on large system limit model. One of the
advantage of RIP based analysis is its generality for any type of
sensing matrices. However,  our large system analysis can provide
explicit bounds for the number of required sensor elements and SNR
requirement thanks to the Gaussian nature of sensing matrix.

\subsection{Comparison with results of Davies and Eldar \cite{DaviesEldar2010} }

Another recent development in joint sparse recovery approaches is
the rank-awareness algorithm by Davies and Eldar
\cite{DaviesEldar2010}. The algorithm is derived in the noiseless
measurement setup and is basically the same as our subspace S-OMP in
Section 5 except that $\ab_j$ in step 3 is normalized after applying
$R^\perp_{R(A_{I_t})}$ to the original dictionary $A$. For the full
rank measurement, i.e. $r=k$, the performance of the rank-aware
subspace S-OMP is equivalent to that of MUSIC. However, for $r<k$,
the lack of the generalized MUSIC criterion may make the algorithm
inferior since our generalized MUSIC criterion can identify $r$
support deterministically whereas the rank-aware subspace S-OMP
should estimate the remaining $r$ support with additional
error-prone greedy steps.

\subsection{Compressive MUSIC with  a mixed norm approach}

Another important issue in CS-MUSIC is how to combine the general
MUSIC criterion with non-greedy joint sparse recovery algorithms
such as a mixed norm approach \cite{malioutov2005ssr}. Towards this, the $k-r$
greedy step required for the analysis for CS-MUSIC should be
modified. One solution to mitigate this problem is to choose a $k-r$ support from
the non-zero support of the solution and use it as a partial support for
generalized MUSIC criterion. However, we still need a
criterion to identify a correct $k-r$ support from the solution, since the generalized MUSIC criterion only holds
with a correct $k-r$ support. Recently, we showed that
a correct $k-r$ partial support out of $k$-sparse solution can be identified using a subspace fitting criterion \cite{KimLeeYeOptimizedSupport}. Accordingly, the joint sparse recovery problem can be relaxed to a problem to find a solution that has at least $k-r+1$ correct support out of $k$ nonzero support estimate. This is a significant relaxation of  CS-MUSIC in its present form that requires  $k-r$ successful consecutive greedy steps. Accordingly, the new formulation was shown to significantly improve the performance of  CS-MUSIC for the joint-sparse recovery \cite{KimLeeYeOptimizedSupport}.
However, the new results are beyond scope of this paper and will be reported separately.

\subsection{Relation with distributed compressive sensing coding region}

Our theoretical results as well as numerical experiments  indicate
that the number of resolvable sources can increase thanks to the
exploitation of the noise subspace.  This observation leads us to
investigate whether  CS-MUSIC  achieves the rate region  in
distributed compressed sensing \cite{baron2005distributed}, which is
analogous to Slepian-Wolf coding regions in distributed source
coding \cite{slepian1973noiseless}.

\begin{figure}[htbp]
 \centerline{\epsfig{figure=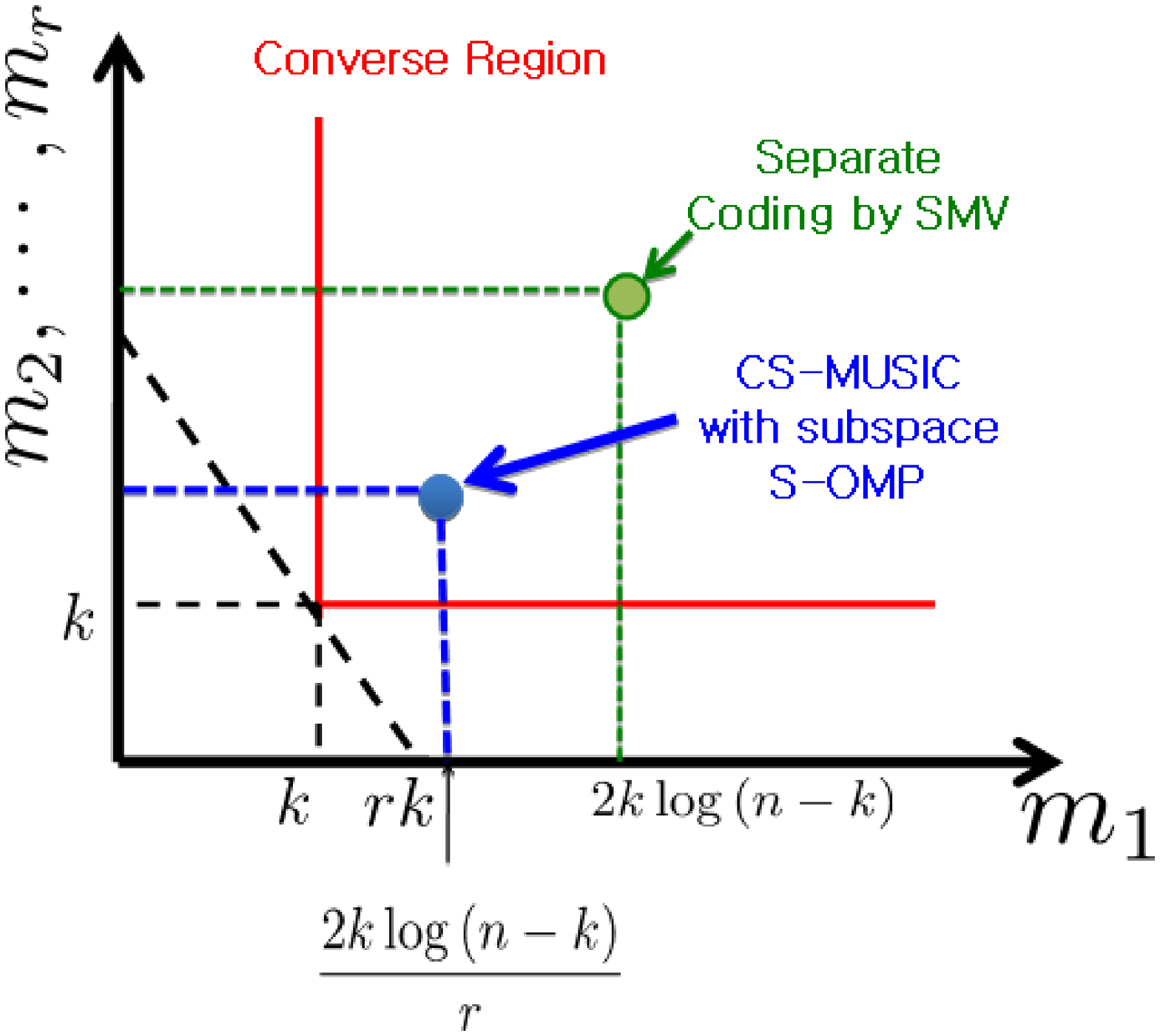,width=8cm}\epsfig{figure=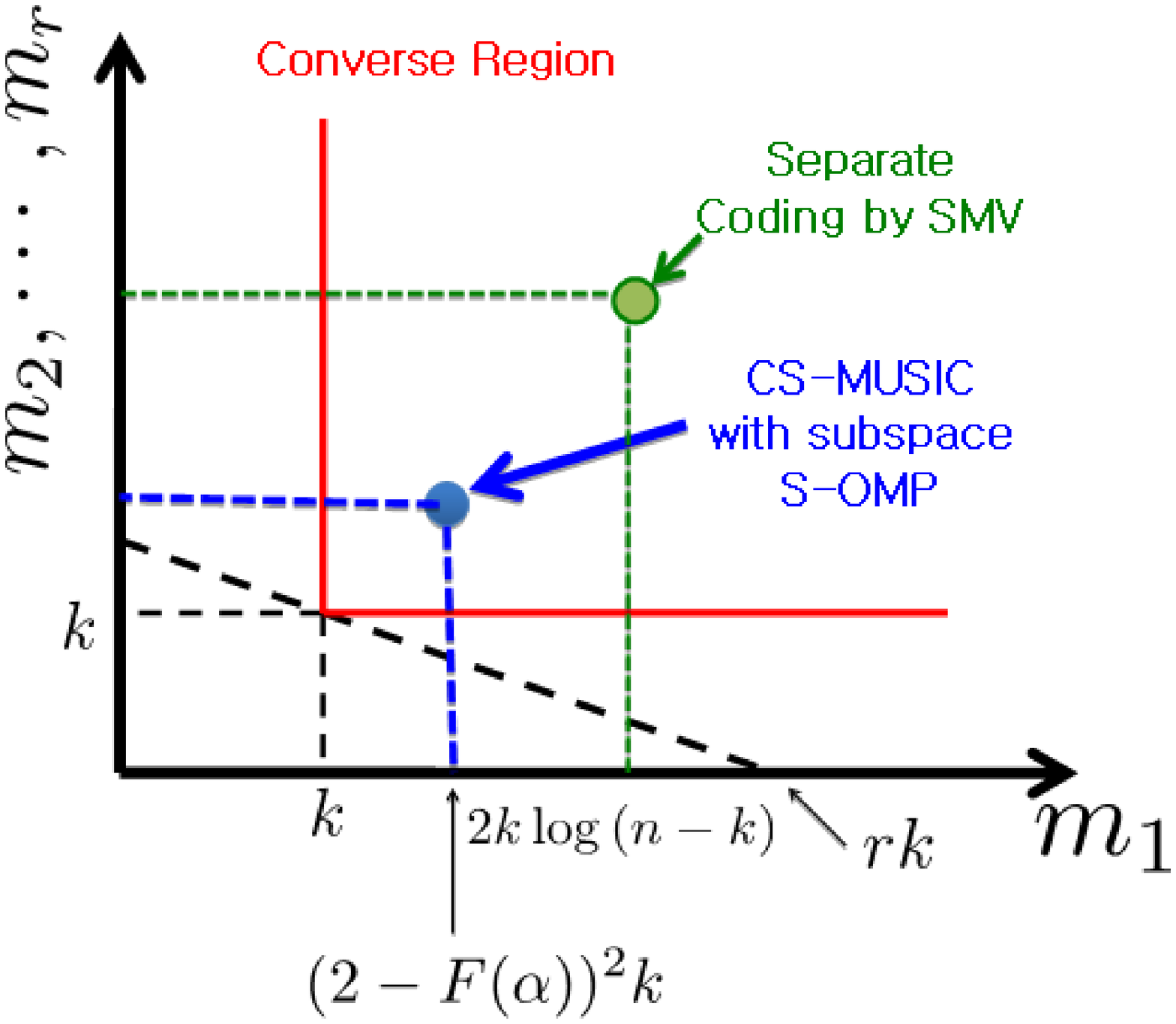,width=8cm}}
 \centerline{ \mbox{(a)}\hspace{8cm}\mbox{(b)}  }
  \caption{Rate regions for the multiple measurement vector problem and CS-MUSIC, when (a) $r$ is a fixed number, and (b) $\lim_{n\rightarrow \infty} r/k=\alpha>0$.}
 \label{fig:coding_region}
\end{figure}

Recall that the necessary condition for a maximum likelihood for SMV
sparse recovery is given by \cite{fletcher2009necessary}: $$ m >
\frac{2 k\log(n-k)}{\sf{SNR}\cdot \sf{MSR}_{\min}^k}+k-1
\longrightarrow k-1,$$ as $\sf{SNR}\rightarrow \infty$. Let $m_i$
denote the number of sensor elements at the $i$-th measurement
vector. If the total number of samples from $r$ vectors are smaller
than that of SMV-CS, i.e. $\sum_{i=1}^r m_i < k-1$, then we cannot
expect a perfect recovery even from noiseless measurement vectors.
Furthermore,  the minimum sensor elements should be $m_i=k$ to
recover the values of the $i$-th coefficient vector, even when the
$k$ indices of the support are correctly identified. Hence, the
converse region at $\textrm{SNR}\rightarrow \infty$ is defined by
the $m_i< k, i=1,\cdots, r$ as shown
Fig.~\ref{fig:coding_region}(a)(b).

Now, for a fixed $r$ our analysis shows that the achievable rate by
the CS-MUSIC is $m_i = 2k\log(n-k)/r$
(Fig.~\ref{fig:coding_region}(a)). On the other hand,  if
$\lim_{n\rightarrow \infty} r/k=\alpha>0$, the achievable rate by
the CS-MUSIC is  $m_i = (2-F(\alpha))^2k$ as shown in
Fig.~\ref{fig:coding_region}(b). Therefore, CS-MUSIC approaches the
converse regin at $r=k$, whereas for the intermediate ranges of $r$
there exists a performance gap from the converse region. However,
even in this case if we consider a separate SMV decoding without
considering correlation structure in MMV,  the required sampling
rate is $m_i\geq 2k\log(n-k)$ which is significantly larger than
that of CS-MUSIC. This analysis clearly reveals that CS-MUSIC is a
quite efficient decoding method from  distributed compressed sensing
perspective.

%
%
%
%

\subsection{Discretization}

The MUSIC algorithm was originally developed for spectral estimation or direction-of-arrival (DOA) estimation problem,  where the unknown target locations and bearing angle are continuously varying parameters.  If we apply CS-MUSIC to this type of problems to achieve a finer resolution, the search region should be discretized more finely with a large $n$.  The main problem of such discretization is that the mutual coherence of the dictionary $A$ approaches to 1, which can violate the RIP condition of the CS-MUSIC.  Therefore, the trade-off between the resolution and the RIP condition should be investigated;
 Duarte and Baraniuk recently investigated such trade-off in the context of spectral compressive sensing \cite{DuarteSpectral}.
 Since this problem is very important  not only for the CS-MUSIC but for SMV compressed sensing   problems that are originated from discretizing  continuous problems, systematic study needs to be done in the future.

\section{Conclusions and future works}\label{sec:con}
\label{sec:conclusion}

In this paper, we developed a novel compressive MUSIC algorithm that
outperforms the conventional MMV algorithms. The algorithm estimates
$k-r$ entries of the support using conventional MMV algorithms,
while the remaining $r$ support indices are
estimated using a generalized MUSIC criterion, which was derived
from the RIP properties of sensing matrix. Theoretical analysis as
well as numerical simulation demonstrated that our compressive MUSIC
algorithm achieved the $l_0$ bound as $r$ approaches the non-zero
support size $k$. This is fundamentally different from existing
information theoretic analysis \cite{tang2009performance}, which
requires the number of snapshots to  approach infinity to achieve
the $l_0$ bound. Furthermore, as $r$ approaches 1, the recovery rate
approaches that of the conventional SMV compressive sensing. We also
provided a method that can estimate the unknown sparsity, even under
noisy measurements. Theoretical analysis based on a large system MMV
model showed that the required number of sensor elements for
compressive MUSIC is much smaller than that of conventional MMV
compressive sensing. Furthermore,  we provided a closed form
expression of the minimum SNR to guarantee the success of compressive
MUSIC.

The compressive sensing and array signal processing produce two
extreme approaches for the MMV problem: one is based on a probabilistic
guarantee, the other on a deterministic guarantee. One
important contribution of this paper is to abandon such extreme viewpoints
and propose an optimal method to take the best of both worlds.
Even though the resulting idea appears simple, we believe that this
opens a new area of research.  Since extensive research results are
available from the array signal processing community,
combining the already well-established results with compressive
sensing may produce algorithms that may be superior to the
compressive MUSIC algorithm in its present form. Another interesting
observation is that the RIP condition $\delta^L_{2k-r+1}<1$, which
is essential for compressive MUSIC to achieve the $l_0$ bound, is
identical to the $l_0$ recovery condition for the so-called modified CS
\cite{vaswani2009modified}. In modified CS, $r$ support indices are
known {\em a priori} and the remaining $k-r$ are estimated using SMV
compressive sensing. The duality between compressive MUSIC and the
modified CS does not appear incidental and should be investigated.
Rather than estimating $k-r$ indices first using MMV compressive
sensing and estimating the remaining $r$ using the generalized MUSIC
criterion, there might be a new algorithm that estimates $r$
supports indices first in a deterministic pattern, while the remaining
$k-r$ are estimated using compressive sensing.  This direction of
research might reveal new insights about the geometry of the MMV
problem.

\section*{Appendix A: Proof of Theorem~\ref{lemspark}}
 \renewcommand{\theequation}{A.\arabic{equation}}
\renewcommand{\thetheorem}{A.\arabic{theorem}}
\renewcommand{\thelemma}{A.\arabic{lemma}}
\setcounter{equation}{0} \setcounter{theorem}{0}
\begin{proof}
\indent (a) First, we show that ${\rm spark}(Q^{*}A)\leq k-r+1$.  Since $Q^{*}AX=0$,  we have
$Q^{*}A\mathbf{x}_i=0$ for $1\leq i\leq r$. Take a set
$P\subset {\rm supp}X$ with $|P|=r-1$. Then, there exists a
nonzero $\mathbf{c}=[c_1,\cdots,c_r]\in\mathbb{R}^r$ such that
\begin{equation}\label{eq-lemspark1}
X^P\mathbf{c}=0,~{\rm where}~X^P\in\mathbb{R}^{(r-1)\times r},
\end{equation}
where $X^P$ denotes a submatrix collecting rows corresponding to the
index set $P$.

 Since the columns of $X$ are linearly independent,
$\sum_{i=1}^r c_i\mathbf{x}_i\neq 0$ but $Q^{*}A(\sum_{i=1}^r
c_i\mathbf{x}_i)= 0$. By (\ref{eq-lemspark1}),
$$\|\sum\limits_{i=1}^rc_i\mathbf{x}_i\|_0\leq k-r+1$$
so that ${\rm spark}(Q^{*}A)\leq k-r+1$. \\
\indent (b) Suppose that there is $\mathbf{x}\in
\mathbb{R}^n\setminus \{\mathbf{0}\}$ such that
$$Q^{*}A\mathbf{x}= 0,~ \|\mathbf{x}\|_0\leq k-r+1~{\rm and}~{\rm
supp}\mathbf{(x)}\nsubseteq {\rm supp}X.$$ Since
$Q^{*}A\mathbf{x}=0$, $A\mathbf{x}\in R(Q)^{\perp}=R(B)$ so that
there is a $\tilde{\mathbf{x}}$ such that
$A\mathbf{x}=A\tilde{\mathbf{x}}$ and ${\rm
supp}(\tilde{\mathbf{x}})\subset {\rm supp}X$. Hence, we have
$$A(\mathbf{x}-\tilde{\mathbf{x}})=0,~\|\mathbf{x}-\tilde{\mathbf{x}}\|_0\leq 2k-r+1.$$
By the RIP condition $0\leq \delta^L_{2k-r+1}(A)<1$, $\xb=\tilde{\xb}$.
It follows that whenever $\|\xb\|_0\leq k-r+1$ and $Q^{*}A\xb=0$, we have ${\rm supp}(\xb)\subset {\rm supp}X$.
Since $A\mathbf{x}\in R(B)=R(AX)$, there is a $\mathbf{y}\in R(X)$
such that $A\mathbf{x}=A\mathbf{y}$. Hence, if $Q^{*}A\mathbf{x}=0$
and $\|\mathbf{x}\|_0\leq k-r+1$, by the RIP condition of $A$, we
have $\mathbf{x}\in R(X)$.

Finally, it suffices to show that  for any $\mathbf{x}\in
R(X)\setminus \{0\}$, $$\|\mathbf{x}\|_0\geq k-r+1.$$ Suppose that
$\|\mathbf{x}\|_0\leq k-r$. Then there is a set $Z$ such that
$|Z|=r$ and $Z\subset {\rm supp}X\setminus {\rm supp}(\xb)$. Then,
there exists a $\mathbf{c}\in\mathbb{R}^r\setminus \{0\}$ such that
\begin{equation*}
X^Z\mathbf{c}=0,~{\rm where}~X^Z\in\mathbb{R}^{r\times r}.
\end{equation*}
This is impossible since the nonzero rows of $X$ are in general
position.

\end{proof}

\section*{Appendix B: Proof of Theorem~\ref{com-music} and Corollary~\ref{coro-comusic}}
 \renewcommand{\theequation}{B.\arabic{equation}}
\renewcommand{\thetheorem}{B.\arabic{theorem}}
\renewcommand{\thelemma}{B.\arabic{lemma}}
\setcounter{equation}{0} \setcounter{theorem}{0}
\indent

{\em Proof of Theorem \ref{com-music}:~}
In order to show that \eqref{eq-comusic} implies $j\in {\rm supp}X$, let $I_{k-r}$ be an index set with $|I_{k-r}|=k-r$ and $I_{k-r}\subset {\rm supp}X$. Then by Lemma 4.1 and the definition of the spark(A),
$${\rm rank}[Q^{*}A_{I_{k-r}}]=k-r.$$ By the assumption, there is an $\xb_{k-r}\in\mathbb{R}^{k-r}$ such that $Q^{*}\ab_j=Q^{*}A_{I_{k-r}}\xb_{k-r}$ so that we have
$$Q^{*}[\ab_j-A_{I_{k-r}}\xb_{k-r}]=0.$$
Since $\ab_j-A_{I_{k-r}}\xb_{k-r}\in N(Q^{*})=R(Q)^{\perp}=R(B)$, there is a $\tilde{\xb}\in\mathbb{R}^n$ such that ${\rm supp}(\tilde{\xb})\subset {\rm supp}X$ and
$\ab_j-A_{I_{k-r}}\xb_{k-r}=A\tilde{\xb}$. Hence we have $\yb\in\mathbb{R}^n$ such that $A\yb=0$ and ${\rm supp}(\yb)\subset {\rm supp}X\cup \{j\}\cup I_{k-r}$ so that $\|\yb\|\leq 2k-r+1$. By the RIP condition $0\leq \delta_{2k-r+1}^L(A)<1$, it follows that
$\{j\}\cup {\rm supp}(\xb_{k-r})={\rm supp}(\tilde{\xb})\subset {\rm supp}X$ since $j\notin I_{k-r}$. Hence, under the condition \eqref{eq-comusic}, we have $j\in {\rm supp}X$.

In order to show that $j\in {\rm supp}X$ implies \eqref{eq-comusic}, assume the contrary. Then we have
$${\rm rank}(Q^{*}[A_{I_{k-r}},\ab_j])=k-r+1,$$
where $I_{k-r}\subset {\rm supp}X$ with $|I_{k-r}|=k-r$.
Then for any $\xb_{k-r}\in\mathbb{R}^{k-r}$, $Q^{*}[\ab_j-A_{I_{k-r}}\xb_{k-r}]\neq 0$ so that $\ab_j-A_{I_{k-r}}\xb_{k-r}\notin R(B)$. Set $P={\rm supp}X\setminus (I_{k-r}\cup \{j\})$ so that $|P|=r-1$. Then there is a $\cb\in\mathbb{R}^r\setminus \{0\}$ such that
$$X^P\cb=0,~{\rm where~}X^P\in\mathbb{R}^{(r-1)\times r}.$$
Then we have $\|X\cb\|_0=k-r+1$ since the rows of $X$ are in general position. Note that ${\rm supp}(X\cb)=\{j\}\cup I_{k-r}$. Since $AX\cb\in R(B)$, $\ab_j-A_{I_{k-r}}\xb_{k-r}\in R(B)$ for some $\xb_{k-r}\in \mathbb{R}^{k-r}$, which is a contradiction.

\bigskip

{\em Proof of Corollary \ref{coro-comusic}:~}
Here we let $G_{I_{k-r}}:=Q^{*}A_{I_{k-r}}$ and $\gb_j=Q^{*}\ab_j$. Since we already have ${\rm rank}[G_{I_{k-r}}]=k-r$, (\ref{eq-comusic}) holds if and only if
$$\det{[G_{I_{k-r}},\gb_j]^{*}[G_{I_{k-r}},\gb_j]}=0.$$ Note that
\begin{eqnarray*}
\left[
\begin{array}{c}
G_{I_{k-r}}^{*}\\ \gb_j^{*}
\end{array}
\right]
[G_{I_{k-r}},\gb_j]&=&\left[
\begin{array}{c}
A_{I_{k-r}}^{*}\\ \ab_j^{*}
\end{array}
\right]QQ^{*}[A_{I_{k-r}},\ab_j]\\
&=&
\left[
\begin{array}{cc}
A_{I_{k-r}}^{*}P_{R(Q)}A_{I_{k-r}} & A_{I_{k-r}}^{*}P_{R(Q)}\ab_j\\
\ab_j^{*}P_{R(Q)}A_{I_{k-r}} & \ab_j^{*}P_{R(Q)}\ab_j
\end{array}
\right],
\end{eqnarray*}
where $\det{[A_{I_{k-r}}^{*}P_{R(Q)}A_{I_{k-r}}]}>0$ because of
${\rm rank}[G_{I_{k-r}}]=k-r$. Since
\begin{eqnarray*}
&&\det\left[
\begin{array}{cc}
A_{I_{k-r}}^{*}P_{R(Q)}A_{I_{k-r}} & A_{I_{k-r}}^{*}P_{R(Q)}\ab_j\\
\ab_j^{*}P_{R(Q)}A_{I_{k-r}} & \ab_j^{*}P_{R(Q)}\ab_j
\end{array}
\right]\\
&=&\det(A_{I_{k-r}}^{*}P_{R(Q)}A_{I_{k-r}})\det(\ab_j^{*}P_{R(Q)}\ab_j-\ab_j^{*}P_{R(Q)}A_{I_{k-r}}
(A_{I_{k-r}}^{*}P_{R(Q)}A_{I_{k-r}})^{-1}A_{I_{k-r}}^{*}P_{R(Q)}\ab_j),
\end{eqnarray*}
(\ref{eq-comusic}) is equivalent to
\begin{eqnarray*}
0&=&\ab_j^{*}P_{R(Q)}\ab_j-\ab_j^{*}P_{R(Q)}A_{I_{k-r}}
(A_{I_{k-r}}^{*}P_{R(Q)}A_{I_{k-r}})^{-1}A_{I_{k-r}}^{*}P_{R(Q)}\ab_j \\
&=&\ab_j^{*}QQ^{*}\ab_j-\ab_j^{*}QP_{R(Q^{*}A_{I_{k-r}})}Q^{*}\ab_j\\
&=&\ab_j^{*}\left[P_{R(Q)}-P_{R(P_{R(Q)}A_{I_{k-r}})}\right]\ab_j,
\end{eqnarray*}
where $P_{R(Q)}=QQ^{*}.$
Hence (\ref{eq-comusicmod}) holds if and only if $j\in {\rm supp}X$.

\section*{Appendix C: Proof of Theorem~\ref{cmusic-geo} and Lemma \ref{lem-supp-est}}
 \renewcommand{\theequation}{C.\arabic{equation}}
\renewcommand{\thetheorem}{C.\arabic{theorem}}
\renewcommand{\thelemma}{C.\arabic{lemma}}
\setcounter{equation}{0} \setcounter{theorem}{0}
\indent

{\em Proof of Theorem \ref{cmusic-geo}:}~
(a) By the definitions of $U$ and $Q$, we have $U^{*}Q=0$ so that
\begin{eqnarray*}
[UU^{*}+P_{QQ^{*}A_{I_{k-r}}}]^2&=&UU^{*}+UU^{*}QQ^{*}A_{I_{k-r}}(A_{I_{k-r}}^{*}QQ^{*}A_{I_{k-r}})^{-1}A_{I_{k-r}}^{*}QQ^{*}\\
&&+QQ^{*}A_{I_{k-r}}(A_{I_{k-r}}^{*}QQ^{*}A_{I_{k-r}})^{-1}A_{I_{k-r}}^{*}QQ^{*}UU^{*}\\
&&+QQ^{*}A_{I_{k-r}}(A_{I_{k-r}}^{*}QQ^{*}A_{I_{k-r}})^{-1}A_{I_{k-r}}^{*}QQ^{*}A_{I_{k-r}}(A_{I_{k-r}}^{*}QQ^{*}A_{I_{k-r}})^{-1}
A_{I_{k-r}}^{*}QQ^{*}\\
&=&UU^{*}+P_{QQ^{*}A_{I_{k-r}}}.
\end{eqnarray*}
Since $UU^{*}+P_{QQ^{*}A_{I_{k-r}}}$ is a self-adjoint matrix, it is an orthogonal projection.
Next, to show that $R(UU^{*}+P_{QQ^{*}A_{I_{k-r}}})=R(B)+R(QQ^{*}A_{I_{k-r}})$, we only need to show the following properties :
\begin{itemize}
\item [(i)] $[UU^{*}+P_{QQ^{*}A_{I_{k-r}}}]\bb=\bb$ for any $\bb\in R(B)$,
\item [(ii)] $[UU^{*}+P_{QQ^{*}A_{I_{k-r}}}]\qb_1=\qb_1$ for any $\qb_1\in R(QQ^{*}A_{I_{k-r}})$,
\item [(iii)] $[UU^{*}+P_{QQ^{*}A_{I_{k-r}}}]\qb_2={\bf 0}$ for any $\qb_2\in R(Q)\cap R(QQ^{*}A_{I_{k-r}})^{\perp}.$
\end{itemize}
For (i), it can be easily shown by using $Q^{*}\bb={\bf 0}$ and $UU^{*}\bb=\bb$ for any $\bb\in R(B)$. For (ii), any
$\qb_1\in R(QQ^{*}A_{I_{k-r}})$, there is a $\wb\in \mathbb{R}^{k-r}$ such that $\qb_1=QQ^{*}A_{I_{k-r}}\wb$. Then by using the property $U^{*}Q=0$, we can see that property (ii) also holds. Finally, we can easily see that (iii) also holds by using $U^{*}\qb=0$ for any $\qb\in R(Q)$.

(b) This is a simple consequence of (a) since $QQ^{*}-P_{QQ^{*}A_{I_{k-r}}}=I-[UU^{*}+P_{QQ^{*}A_{I_{k-r}}}]$ and
$R(Q)\cap R(QQ^{*}A_{I_{k-r}})^{\perp}$ is an orthogonal complement of $R(B)+R(QQ^{*}A_{I_{k-r}})$.

(c) Since ${\rm spark}(Q^{*}A)=k-r+1$ by Lemma \ref{lemspark},
$[U~~A_{I_{k-r}}]$ has $k$ linearly independent columns. Hence we
only need to find the orthogonal complement of
$R([U~~A_{I_{k-r}}])=R(U)+R(A_{I_{k-r}})=R(B)+R(A_{I_{k-r}})$. Since
$R(U)^{\perp}=R(Q)$, we have
$R(U)+R(A_{I_{k-r}})=R(U)+R(P_QA_{I_{k-r}})$ by the projection
update rule so that $(R(U)+R(QQ^{*}A_{I_{k-r}}))^{\perp}=R(Q)\cap
R(QQ^{*}A_{I_{k-r}})^{\perp}$ is the noise subspace for
$[U~~A_{I_{k-r}}]$ or $[B~~A_{I_{k-r}}]$.

\section*{Appendix D: Proof of Theorem~\ref{genmusic-noisy}}
 \renewcommand{\theequation}{D.\arabic{equation}}
\renewcommand{\thetheorem}{D.\arabic{theorem}}
\renewcommand{\thelemma}{D.\arabic{lemma}}
\setcounter{equation}{0} \setcounter{theorem}{0}
\indent
We first need to show the following lemmas.
\begin{lemma}\label{lem:cons}
Assume that we have noisy measurement through multiple noisy
snapshots where
$$Y=AX+N,$$
where $A\in\mathbb{R}^{m\times n}$, $X\in\mathbb{R}^{n\times r}$,
and $N\in\mathbb{R}^{m\times r}$ is additive noise. We also
assume that $I_{k-r}\subset {\rm supp}X$. Then there is a $\eta>0$
such that for any $j\notin {\rm supp}X$ and $l\in {\rm supp}X$,
\begin{equation}\label{thm-cons-cond}
\ab_j^{*}\left[P_{R(\hat{Q})}-P_{R(P_{R(\hat{Q})}A_{I_{k-r}})}\right]\ab_j>\ab_l^{*}\left[P_{R(\hat{Q})}-P_{R(P_{R(\hat{Q})}A_{I_{k-r}})}\right]\ab_l
\end{equation}
if $\|N\|<\eta$, where $\|N\|$ is a spectral norm of $N$ and $\hat{Q}\in\mathbb{R}^{m\times (m-k)}$ consists of orthonormal columns such that $\hat{Q}^{*}Y=0$.
\end{lemma}
\begin{proof}
First, here we let $B=AX$, $\sigma_{\min}(B)$ (or $\sigma_{\max}(B)$)
be the minimum (or the maximum) nonzero singular value of $B$. Then,
$Y=B+N$ is also of full column rank if $\|N\|<\sigma_{\min}(B)$. For
such an $N$,
\begin{eqnarray*}
&&\|P_{R(Y)}-P_{R(B)}\|=\|Y(Y^{*}Y)^{-1}Y^{*}-B(B^{*}B)^{-1}B^{*}\|\\
&=&\|(B+N)[(B+N)^{*}(B+N)]^{-1}(B+N)^*-B(B^{*}B)^{-1}B^{*}\|\\
&\leq& \|N\|\|[(B+N)^{*}(B+N)]^{-1}(B+N)^*\|\\
&&+\|(B+N)[(B+N)^{*}(B+N)]^{-1}\|\|(B+N)^{*}(B+N)-B^{*}B\|\|(B^{*}B)^{-1}B^{*}\|
+\|B(B^{*}B)^{-1}\|\|N\|\\
&\leq&\|N\|(B+N)^{\dagger}\|+\|(B+N)^{\dagger}\|\|B^{\dagger}\|[2\|B\|\|N\|+\|N\|^2]+\|B^{\dagger}\|\|N\|
\end{eqnarray*}
by the consecutive use of triangle inequality. If we have $\|N\|<\sigma_{\min}(B)$, we get $$\|(B+N)^{\dagger}\|\leq (\sigma_{\min}(B)-\|N\|)^{-1}$$ so that
\begin{eqnarray}\label{cons-eq1}
\notag \frac{\|P_{R(Y)}-P_{R(B)}\|}{\|Y-B\|}&\leq&
\frac{1}{\sigma_{\min}(B)-\|N\|}+\frac{1}{\sigma_{\min}(B)(\sigma_{\min}(B)-\|N\|)}\left[2\|B\|+\|N\|\right]+\frac{1}{\sigma_{\min}(B)}\\
&=&\frac{2(\sigma_{\max}(B)+\sigma_{\min}(B))}{\sigma_{\min}(B)(\sigma_{\min}(B)-\|N\|)}
\end{eqnarray}
where we use $\|B^{\dagger}\|=1/(\sigma_{\min}(B))$ and
$\|B\|=\sigma_{\max}(B)$. By the projection update rule, we have
\begin{eqnarray}\label{aa-proj-b}
P_{R([B~~A_{I_{k-r}}])}&=&P_{R(B)}+P_{R(P_{R(B)}^{\perp}A_{I_{k-r}})}=I-\left[P_{R(Q)}-P_{R(P_{R(Q)}A_{I_{k-r}})}\right]\\
\notag &=&P_{R(A_{I_{k-r}})}+P_{R(P_{R(A_{I_{k-r}})}^{\perp}B)},
\end{eqnarray}
and similarly,
\begin{eqnarray}\label{aa-proj-y}
P_{R([Y~~A_{I_{k-r}}])}&=&P_{R(Y)}+P_{R(P_{R(Y)}^{\perp}A_{I_{k-r}})}=I-\left[P_{R(\hat{Q})}-P_{R(P_{R(\hat{Q})}A_{I_{k-r}})}\right]\\
\notag &=&P_{R(A_{I_{k-r}})}+P_{R(P_{R(A_{I_{k-r}})}^{\perp}Y)}.
\end{eqnarray}
By applying \eqref{aa-proj-b} and \eqref{aa-proj-y} as done in \cite{LeeBresler2010}, we have
\begin{eqnarray}\label{aa-pert}
\notag \|[P_{R(\hat{Q})}-P_{R(P_{R(\hat{Q})}A_{I_{k-r}})}]-[P_{R(Q)}-P_{R(P_{R(Q)}A_{I_{k-r}})}]\|&=&\|
P_{R(P_{R(A_{I_{k-r}})}^{\perp}Y)}-
{P_{R(P_{R(A_{I_{k-r}})}^{\perp}B)}}\|\\
&\leq&\|P_{R(Y)}-P_{R(B)}\|.
\end{eqnarray}
Then, for any $j\notin {\rm supp}X$ and $l\in {\rm supp}X$, by the generalized MUSIC criterion \eqref{eq-comusicmod} we have
\begin{eqnarray*}
&&\ab_j^{*}\left[P_{R(\hat{Q})}-P_{R(P_{R(\hat{Q})}A_{I_{k-r}})}\right]\ab_j-\ab_l^{*}\left[P_{R(\hat{Q})}-P_{R(P_{R(\hat{Q})}A_{I_{k-r}})}\right]\ab_l\\
&=&\ab_j^{*}\left[P_{R(Q)}-P_{R(P_{R(Q)}A_{I_{k-r}})}\right]\ab_j-\ab_l^{*}\left[P_{R(Q)}-P_{R(P_{R(Q)}A_{I_{k-r}})}\right]\ab_l\\
&&+\ab_j^{*}\left[[P_{R(\hat{Q})}-P_{R(P_{R(\hat{Q})}A_{I_{k-r}})}]-[P_{R(Q)}-P_{R(P_{R(Q)}A_{I_{k-r}})}]\right]\ab_j\\
&&-\ab_l^{*}\left[[P_{R(\hat{Q})}-P_{R(P_{R(\hat{Q})}A_{I_{k-r}})}]-[P_{R(Q)}-P_{R(P_{R(Q)}A_{I_{k-r}})}]\right]\ab_l\\
&\geq&\min\limits_{j\notin{\rm supp}X}\ab_j^{*}\left[P_{R(\hat{Q})}-P_{R(P_{R(\hat{Q})}A_{I_{k-r}})}\right]\ab_j-2\max(\|\ab_j\|^2,\|\ab_l\|^2)\|P_{R(Y)}-P_{R(B)}\|\\
&\geq&\min\limits_{j\notin{\rm supp}X}\ab_j^{*}\left[P_{R(\hat{Q})}-P_{R(P_{R(\hat{Q})}A_{I_{k-r}})}\right]\ab_j-2\max\limits_{1\leq j\leq n}\|\ab_j\|^2
\frac{2(\sigma_{\max}(B)+\sigma_{\min}(B))\|N\|}{\sigma_{\min}(B)(\sigma_{\min}(B)-\|N\|)}>0
\end{eqnarray*}
if we have
\begin{equation}\label{noise-bound}
\|N\|<\frac{\sigma_{\min}^2(B)\zeta}{4(\sigma_{\max}(B)+\sigma_{\min}(B))+\sigma_{\min}(B)\zeta}
\end{equation}
where
$$\zeta:=\frac{\min\limits_{j\notin{\rm supp}X}\ab_j^{*}\left[P_{R(\hat{Q})}-P_{R(P_{R(\hat{Q})}A_{I_{k-r}})}\right]\ab_j}{\max\limits_{1\leq j\leq n}\|\ab_j\|^2}.$$
\end{proof}

\begin{lemma}\label{lemma-snr}
Suppose a minimum $\textsf{SNR}$ is given by
$$\textsf{SNR}_{\min}(Y):=\frac{\sigma_{\min}(B)}{\|N\|}\geq \eta,$$
where
\begin{equation*}
\eta:=1+\frac{4(\kappa(B)+1)}{\zeta},
\end{equation*}
$\kappa(B)$ is the condition number of
$B=AX$ and
$$\zeta:=\frac{\min\limits_{j\notin{\rm supp}X}\ab_j^{*}\left[P_{R(\hat{Q})}-P_{R(P_{R(\hat{Q})}A_{I_{k-r}})}\right]\ab_j}{\max\limits_{1\leq j\leq n}\|\ab_j\|^2}.$$ Then, for any $j\notin {\rm supp}X$ and $l\in {\rm supp}X$,
$$\ab_j^{*}\left[P_{R(\hat{Q})}-P_{R(P_{R(\hat{Q})}A_{I_{k-r}})}\right]\ab_j>\ab_l^{*}\left[P_{R(\hat{Q})}-P_{R(P_{R(\hat{Q})}A_{I_{k-r}})}\right]\ab_l.$$
\end{lemma}
\begin{proof}
Using \eqref{noise-bound}, the generalized MUSIC correctly estimates the $r$
remaining indices when
\begin{equation*}
\|N\| < \frac{\sigma_{\min}(B)\zeta}{4(\kappa(B)+1)+\zeta}
\end{equation*}
where we use the definition of the condition number of the $B=AX$
matrix, i.e.  $\kappa(AX) =\frac{\sigma_{\max}(B)}{\sigma_{\min}(B)}$.
This implies that
\begin{equation*}
\textsf{SNR}_{\min}(Y) >
1+\frac{4(\kappa(B)+1)}{\zeta}.
\end{equation*}
This concludes the proof.
\end{proof}

\begin{corollary}
For a $\mathrm{LSMMV}(m,n,k,r;\epsilon)$, if we have $I_{k-r}\subset {\rm supp}X$ and  a minimum {\sf SNR} satisfies
\begin{equation}\label{snr-genmusic}
{\sf SNR}_{\min}(Y)> 1+\frac{4(\kappa(B)+1)}{1-\gamma^2} \geq
1+{4(\kappa(B)+1)}
\end{equation}
where $\gamma=\lim_{n\rightarrow\infty}\sqrt{k(n)/m(n)}$, then we can find remaining $r$ indices of ${\rm supp}X$ with generalized MUSIC criterion.
\end{corollary}

\begin{proof}
It is enough to show that
$$\lim\limits_{n\rightarrow\infty}\zeta(n)=1-\gamma^2.$$ First, for each $1\leq j\leq n$, $m\|\ab_j\|^2$ is a chi-square random variable with degree of freedom $m$ so that we have by Lemma 3 in \cite{fletcher2009necessary},
\begin{equation*}
\lim\limits_{n\rightarrow\infty}\frac{\max\limits_{1\leq j\leq n}\|\ab_j\|^2}{m}=1
\end{equation*}
since $\lim_{n\rightarrow\infty}(\log{n})/m=0$.
On the other hand, for any $j\notin {\rm supp}X$, $\ab_j$ is independent from $P_{R(\hat{Q})}-P_{R(P_{R(Q)}A_{I_{k-r}})}$ so that $m\ab_j^{*}\left[P_{R(\hat{Q})}-P_{R(P_{R(\hat{Q})}A_{I_{k-r}})}\right]\ab_j$ is a chi-square random variable with degree of freedom $m-k$ since $P_{R(\hat{Q})}-P_{R(P_{R(\hat{Q})}A_{I_{k-r}})}$ is the projection operator onto the orthogonal complement of $R[B~A_{I_{k-r}}]$. Since $\lim_{n\rightarrow\infty}(\log{(n-k)})/(m-k)=0$, again by Lemma 3 in \cite{fletcher2009necessary}, we have
\begin{equation*}
\lim\limits_{n\rightarrow\infty}\frac{\min\limits_{j\notin{\rm supp}X}\ab_j^{*}\left[P_{R(\hat{Q})}-P_{R(P_{R(\hat{Q})}A_{I_{k-r}})}\right]\ab_j}{m-k}=1
\end{equation*}
so that
\begin{eqnarray*}
&&\lim\limits_{n\rightarrow\infty}\frac{\min\limits_{j\notin {\rm supp}X}\ab_j^{*}
[P_{R(\hat{Q})}-P_{R(P_{R(\hat{Q})}A_{I_{k-r}})}]\ab_j}{\max\limits_{1\leq j\leq n}
\|\ab_j\|^2}\\
&=&\lim\limits_{n\rightarrow\infty}\frac{\min\limits_{j\notin {\rm
supp}X}\ab_j^{*}\left[P_{R(\hat{Q})}-P_{R(P_{R(\hat{Q})}A_{I_{k-r}})}\right]\ab_j}{m-k}\frac{m}{\max\limits_{1\leq
j\leq n}\|\ab_j\|^2}\frac{m-k}{m}=1-\gamma^2 \leq 1.
\end{eqnarray*}
\end{proof}

\bigskip

{\em Proof of Theorem \ref{genmusic-noisy}:}~First, we need to show the left RIP condition $0\leq\delta_{2k-r+1}^L<1$ to apply the generalized MUSIC criterion. Using Mar\'{c}enko-Pastur theorem \cite{marcenko1967distribution}, we have
\begin{equation*}
\limsup\limits_{n\rightarrow\infty}\delta_{2k-r+1}^L=1-\liminf\limits_{n\rightarrow\infty}
(1-\sqrt{(2k-r+1)/m})^2<1.
\end{equation*}
Hence, we need $m\geq (1+\delta)(2k-r+1)$ to make $\limsup_{n\rightarrow\infty}\delta_{2k-r+1}^L>0$. Second, we need to calculate the condition for the number of sensor elements for the SNR condition \eqref{snr-genmusic}. Since $\gamma=\lim_{n\rightarrow\infty}\sqrt{k/m}$, we have \eqref{snr-genmusic} provided that
\begin{equation*}
m\geq k(1+\delta)\left[1-\frac{4(\kappa(B)+1)}{{\sf SNR}_{\min}(Y)-1}\right]^{-1}.
\end{equation*}
Therefore, if  we have ${\sf SNR}_{\min}(Y)>1+{4(\kappa(B)+1)}$ and
$$m\geq \max\left\{k(1+\delta)\left[1-\frac{4(\kappa(B)+1)}{{\sf SNR}_{\min}(Y)-1}\right]^{-1},(1+\delta)(2k-r+1)\right\},$$
then we can identify the remaining $r$ indices of ${\rm supp}X$.


\section*{Appendix E}
\setcounter{equation}{0} \setcounter{theorem}{0} \indent

 \renewcommand{\theequation}{E.\arabic{equation}}
\renewcommand{\thetheorem}{E.\arabic{theorem}}
\renewcommand{\thelemma}{E.\arabic{lemma}}
\setcounter{equation}{0} \setcounter{theorem}{0}

The following two lemmas are quite often used in this paper.

\begin{lemma}\label{chi-max}
Suppose that $r$ is a given number, and $\{u_j^{(n)}\}_{j=1}^n$ is a set of i.i.d. chi-squared random variables with degree of freedom $r$. Then
$$\lim\limits_{n\rightarrow\infty}\max\limits_{j=1,\cdots,n}\frac{u_j^{(n)}}{2\log{n}}=1$$
in probability.
\end{lemma}
\begin{proof}
Assume that $Z_r$ is a chi-squared random variable of degree of $r$, then we have
\begin{equation}\label{gamma-tail}
P\{Z_r>x\}=\frac{\Gamma(r/2,x/2)}{\Gamma(r/2)},
\end{equation}
where $\Gamma(k,z)$ denotes the upper incomplete Gamma function. Then we use the following asymptotic behavior :
$$P\{Z_r>x\}\sim \frac{1}{\Gamma(r/2)}x^{r/2-1}e^{-x/2}
~{\rm as}~ x\rightarrow \infty.$$ For $n\rightarrow\infty$, we
consider the probability $P\{\max_{1\leq j\leq
n}u_j^{(n)}>2(1+\epsilon)\log{n}\}$. By using union bound, we see
that
\begin{eqnarray*}
&&P\left\{\max_{1\leq j\leq n}u_j^{(n)}>2(1+\epsilon)\log{n}\right\}\\
&\leq&n\frac{1}{\Gamma(r/2)}(2(1+\epsilon)\log{n})^{r/2-1}e^{-(1+\epsilon)\log{n}}\\
&\leq&\frac{1}{\Gamma(r/2)}(2(1+\epsilon)\log{n})^{r/2-1}n^{-\epsilon}\rightarrow 0
\end{eqnarray*}
as $n\rightarrow\infty$. Now, considering the probability
$P\{\max_{1\leq j\leq n}u_j^{(n)}<2(1-\epsilon)\log{n}\}$, we see
that
\begin{eqnarray*}
&&P\left\{\max_{1\leq j\leq n}u_j^{(n)}<2(1+\epsilon)\log{n}\right\}\\
&\leq&\left(1-\frac{1}{\Gamma(r/2)}(2(1-\epsilon)\log{n})^{r/2-1}e^{-(1-\epsilon)\log{n}}\right)^n\\
&\leq&\left(1-\frac{1}{\Gamma(r/2)}(2(1-\epsilon)\log{n})^{r/2-1}\frac{1}{n^{1-\epsilon}}\right)^n\rightarrow 0
\end{eqnarray*}
as $n\rightarrow\infty$ so that the claim is proved.
\end{proof}

\begin{lemma}\label{lem:xb-equiv}
Let $A\in\mathbb{R}^{m\times n}$ be the Gaussian sensing matrix whose components $a_{i,j}$ are independent random variable with distribution $\mathcal{N}(0,1/m)$. Then
$$\lim_{n\rightarrow\infty}\frac{\|AX\|_F^2}{\|X\|_F^2}=1.$$
\end{lemma}

\begin{proof}
Because $a_{i,j}\sim \mathcal{N}(0,1/m)$ for all $1\leq i\leq m$ and $1\leq j\leq n$, $m\|\ab_j\|^2$ is a chi-squared random variable of degree of freedom $m$ so that by Lemma 3 in \cite{fletcher2009necessary}, we have
\begin{equation}\label{lem-gaus1}
\lim\limits_{n\rightarrow\infty}\max\limits_{1\leq j\leq n}\|\ab_j\|^2=\lim\limits_{n\rightarrow\infty}\min\limits_{1\leq j\leq n}\|\ab_j\|^2=1.
\end{equation}
Since, for fixed $1\leq j\leq n$, $\ab_i(i\neq j)$ is $m$-dimensional random vector that is nonzero with a probability of 1 and independent of $\ab_j$, the random variable $u(i,j)=\ab_i^{*}\ab_j/\|\ab_j\|$ is a Gaussian random variable with a variance of  $1/m$ by applying Lemma 2 in \cite{fletcher2009necessary}. Since we have \eqref{lem-gaus1} and the variance of $u(i,j)$ goes to 0 as $n\rightarrow\infty$,
\begin{equation}\label{lem-gaus2}
\lim\limits_{n\rightarrow\infty}\ab_i^{*}\ab_j=0
\end{equation}
for all $1\leq j<k\leq n$. Then
\begin{equation*}
\frac{\|AX\|_F^2}{\|X\|_F^2}=\frac{{\rm trace}(X^{*}A^{*}AX)}{{\rm trace}(X^{*}X)}\rightarrow 1
\end{equation*}
as $n\rightarrow\infty$.
\end{proof}

\section*{Appendix F: Proof of Theorem~\ref{lem:num-thres}}
\setcounter{equation}{0} \setcounter{theorem}{0} \indent

 \renewcommand{\theequation}{F.\arabic{equation}}
\renewcommand{\thetheorem}{F.\arabic{theorem}}
\renewcommand{\thelemma}{F.\arabic{lemma}}
\setcounter{equation}{0} \setcounter{theorem}{0}
{\em Proof of Theorem \ref{lem:num-thres}:}
Let $I_t\subset {\rm supp}X$ with $|I_t|=k-r$, where $I_t$ is constructed by the first $k-r$ indices of $X$ if we are ordering the values of $\|\xb^i\|^2$ for $1\leq i\leq n$ with decreasing order. Then for $i\in I_t$,
$$\ab_i^{*}B=\|\ab_i\|^2\xb^i+\ab_i^{*}E^i$$
where $E_i=[\eb_1^i,\cdots,\eb_r^i]\in\mathbb{R}^{n\times r}$ and $\eb_l^i=\bb_l-\ab
_ix_{i,l}.$ Then
\begin{eqnarray}\label{th-supp}
\notag \ab_i^{*}BB^{*}\ab_i&=&\|\|\ab_i\|^2\xb^i+\ab_i^{*}E^i\|^2\\
&\geq&|\|\ab_i\|^2\|\xb^i\|-\|\ab_i^{*}E^i\||^2=\left|\sqrt{A_i}-\sqrt{\sum\limits_{l=1}^{r}B_l^iZ_l^i}\right|^2
\end{eqnarray}
where
\begin{equation*}
A_i=\|\ab_i\|^4\|\xb^i\|^2,~B_l^i=\|\ab_i\|^2\|\eb_l^i\|^2,~Z_l^i=\frac
{|\ab_i^{*}\eb_l^i|^2}{\|\ab_i\|^2\|\eb_l^i\|^2}.
\end{equation*}
First, by Lemma 3 in \cite{fletcher2009necessary}, $\lim_{n\rightarrow\infty}\sup_{i\in I_t}\|\ab_i\|^2=1$ so that we have
\begin{equation}\label{th-a}
\liminf\limits_{n\rightarrow\infty}\frac{A_i}{r{\sf MSR}_{\min}^{k-r}}=\liminf\limits_{n\rightarrow\infty}\frac{\|\ab_i\|^4}{{\sf MSR}_{\min}^{k-r}}
\frac{\|\xb^i\|^2}{r}
\geq 1
\end{equation}
by the definition of ${\sf MSR}_{\min}^{k-r}$ and the construction of $I_t$.

For $B_l^i$, observe that each $\eb_l^i$ is a Gaussian $m$-dimensional vector with total variance
\begin{equation*}
V_l^i:=E[\|\eb_l^i\|^2]\leq E[\|\bb_l\|^2]=\|\xb_l\|^2
\end{equation*}
and $(m/V_l^i)\|\eb_l^i\|^2$ is a chi-squared distribution with a degree of freedom $m$ for $i\in I_t$. Hence using Lemma 3 in \cite{fletcher2009necessary} and
\begin{equation}\label{ineq-thres1}
\frac{\log{(k-r)}}{m}\leq \frac{\log{m}}{m}\longrightarrow 0
\end{equation}
as $n\rightarrow\infty$ so that
\begin{equation*}
\limsup\limits_{n\rightarrow\infty}\max\limits_{i\in I_t}
\frac{\|\ab_i\|^2\|\eb_l^i\|^2}{\|\xb_l\|^2}\leq
\limsup\limits_{n\rightarrow\infty}\max\limits_{i\in I_t}
\frac{\|\ab_i\|^2V_i^l}{\|\xb_l\|^2}\leq 1
\end{equation*}
so that we have
\begin{equation*}
\limsup\limits_{n\rightarrow\infty}\max\limits_{i\in I_t}
\frac{B_l^i}{\|\xb_l\|^2}\leq 1
\end{equation*}
for $i\in I_t$ and $1\leq l\leq r$.
Finally,
$$Z_l^i=\frac{|\ab_i^{*}\eb_l^i|^2}{\|\ab_i\|^2\|\eb_l^i\|^2}$$
follows beta distribution ${\rm Beta}(1,m-1)$ as shown in \cite{fletcher2009necessary}. Since there are $k-r$ terms in $I_t$, Lemma 6 in \cite{fletcher2009necessary} and inequality \eqref{ineq-thres1} shows that
\begin{equation*}
\limsup\limits_{n\rightarrow\infty}\max\limits_{i\in I_t}\frac{m}{2\log{(k-r)}}Z_l^i\leq 1
\end{equation*}
so that we have
\begin{eqnarray}\label{th-bz}
\limsup\limits_{n\rightarrow\infty}\max\limits_{i\in I_t}
\frac{m}{2r\log{(k-r)}}\frac{\sum\limits_{l=1}^r B_l^iZ_l^i}{\|X\|_F^2/r}&\leq&
\sum\limits_{l=1}^r \limsup\limits_{n\rightarrow\infty}\max\limits_{i\in I_t}
\frac{\|\xb_l\|^2}{\|X\|_F^2}\frac{m}{2\log{(k-r)}}Z_l^i\leq 1.
\end{eqnarray}
For $i\notin {\rm supp}X$, we have
\begin{eqnarray*}
\|\ab_i^{*}B\|^2&=&\ab_i^{*}BB^{*}\ab_i=\sum\limits_{l=1}^r \sigma_l^2(B)\|\ab_j^{*}\ub_l\|^2\\
&=&\sigma_{\min}^2(B)\sum\limits_{l=1}^r \|\ab_i^{*}\ub_l\|^2+
\sum\limits_{l=1}^r (\sigma_l^2(B)-\sigma_{\min}^2(B))\|\ab_i^{*}\ub_l\|^2
\end{eqnarray*}
where $B=U\Sigma V$ is the singular value decompostion of $B$,
$U=[\ub_1,\cdots,\ub_l]$ and $\Sigma={\rm
diag}[\sigma_1(B),\cdots,\sigma_r (B)]$ where $\sigma_1(B)\geq
\cdots \geq \sigma_r(B)=\sigma_{\min}(B)>0$. As will be shown later,
the decomposition in the second line of the above equation is
necessary to deal with different asymptotic behavior of chi-square
random variable of degree of freedom 1 and $r$. Since $\ab_i$ is
statistically independent from $\{\ub_l\}_{l=1}^r$ for $i\notin {\rm
supp}X$ and $\{\ub_l\}_{l=1}^r$ is an orthonormal set, $\sum_{l=1}^r
m\|\ab_j^{*}\ub_l\|^2$ is a chi-squared random variable of degree of
freedom $r$ and each $m\|\ab_j^{*}\ub_l\|$ is a chi-squared random
variable of degree of freedom 1. Also, we have
\begin{equation}\label{bnkr-3}
\limsup\limits_{n\rightarrow\infty}\frac{\sum\limits_{l=1}^r(\sigma_l^2(B)-\sigma_{\min}^2(B))m\|\ab_i^{*}\ub_l\|^2}{2r\log{((n-k)r)}}
\leq \limsup\limits_{n\rightarrow\infty}
\left(\frac{\|B\|_F^2}{r}-\sigma_{\min}^2(B)\right)
\end{equation}
since
$$\limsup\limits_{n\rightarrow\infty}\max_{i\notin {\rm supp}X,1\leq l\leq r}
\frac{m\|\ab_j\ub_l\|^2}{2\log{((n-k)r)}}\leq 1
$$
by Lemma 4 in \cite{fletcher2009necessary}.   When $r$ is a fixed number, then by Lemma \ref{chi-max}, we have
\begin{equation}\label{bnkr-1}
\lim\limits_{n\rightarrow\infty}\max\limits_{j=1,\cdots,n}\frac
{\sum\limits_{l=1}^rm\|\ab_j^{*}\ub_l\|^2}{2\log{(n-k)}}=1.
\end{equation}
On the other hand, when $r$ is proportionally increasing with respect to $k$ \cite{fletcher2009necessary}, then we have
\begin{equation}\label{bnkr-2}
\lim\limits_{n\rightarrow\infty}\max\limits_{j=1,\cdots,n}\frac{\sum\limits_{l=1}^rm\|\ab_j^{*}\ub_l\|^2}{r}=1.
\end{equation}
Combining \eqref{bnkr-1}, \eqref{bnkr-2} and \eqref{bnkr-3}, we have
\begin{equation}\label{th-notsupp}
\limsup\limits_{n\rightarrow\infty}\frac{m\|\ab_i^{*}B\|^2}{2B(n,k,r)}\leq 1
\end{equation}
for $j\notin {\rm supp}X$, when $B(n,k,r)$ is given by \eqref{bnkr-def}.

For the noisy measurement $Y$, we have for all $1\leq j\leq n$,
\begin{equation}\label{th-noisy}
\limsup\limits_{n\rightarrow\infty}|\|\ab_j^{*}B\|^2-\|\ab_j^{*}Y\|^2|\leq
\limsup\limits_{n\rightarrow\infty}\|\ab_j\|^2(2\|B\|\|N\|+\|N\|^2)=(2\|B\|+\|N\|)\|N\|.
\end{equation}
Let
\begin{eqnarray*}
\lambda:={\sf MSR}_{\min}^{k-r},&~&\mu:=\frac{\|X\|_F^2}{r}
2\log{(k-r)}\\
\nu:=\frac{2(2\|B\|+\|N\|)\|N\|}{r},&~&\xi=\frac{2B(n,k,r)}{r}.
\end{eqnarray*}
Then, for $i\in {\rm supp}X$, combining \eqref{th-supp}, \eqref{th-a}, \eqref{th-bz} and \eqref{th-noisy}, we have
\begin{eqnarray*}
\liminf\limits_{n\rightarrow\infty}\frac{m\|\ab_i^{*}Y\|^2-m(2\|B\|+\|N\|)\|N\|}{r\xi}&\geq&\liminf\limits_{n\rightarrow\infty}\frac{m\|\ab_i^{*}B\|-2m(2\|B\|+\|N\|)\|N\|}{r\xi}\\
&\geq&\liminf\limits_{n\rightarrow\infty}\frac{\left([\sqrt{\lambda}\sqrt{m}-\sqrt{\mu}]^2-\nu m\right)}{\xi}.
\end{eqnarray*}
On the other hand, using \eqref{th-notsupp} and \eqref{th-noisy}, for $j\notin {\rm supp}X$ we have
\begin{equation*}
\limsup\limits_{n\rightarrow\infty}\frac{m\|\ab_i^{*}Y\|^2-m(2\|B\|+\|N\|)\|N\|}{r\xi}\leq
\limsup\limits_{n\rightarrow\infty}\frac{m\|\ab_i^{*}B\|^2}{r\xi}\leq 1
\end{equation*}
so that we need to show that
\begin{equation}\label{th-success}
\liminf\limits_{n\rightarrow\infty}\frac{[\sqrt{\lambda}\sqrt{m}-\sqrt{\mu}]^2-\nu m}{\xi}
\geq 1+\delta
\end{equation}
under the condition \eqref{snr-thres} and \eqref{num-thres1}. First, note that $\lambda>\nu$
if and only if
\begin{equation*}
r{\sf MSR}_{\min}^{k-r}>2(2\|B\|+\|N\|)\|N\|.
\end{equation*}
which is equivalent to that
\begin{equation*}
\frac{r{\sf MSR}_{\min}^{k-r}}{\sigma_{\min}^2(B)}{\sf SNR}_{\min}^2(Y)-4\kappa(B)
{\sf SNR}_{\min}(Y)-2>0
\end{equation*}
which holds under the condition \eqref{snr-thres}, where we used the
definition
$\kappa(B):=\|B\|/\sigma_{\min}(B)=\sigma_{\max}(B)/\sigma_{\min}(B)$.
Then we can see that if we have $\sqrt{m}\geq
\frac{\sqrt{\mu}}{\sqrt{\lambda}-\sqrt{\nu}},$ then
\begin{equation}\label{th-aux-ineq1}
(\sqrt{\lambda}\sqrt{m}-\sqrt{\mu})^2-\nu m\geq [(\sqrt{\lambda}-\sqrt{\nu})\sqrt{m}-\sqrt{\mu}]^2.
\end{equation}
Also, if we have
$$\sqrt{m}\geq \frac{\sqrt{\mu}+\sqrt{1+\delta}\sqrt{\xi}}{\sqrt{\lambda}-\sqrt{\nu}},$$
then
\begin{equation}\label{th-aux-ineq2}
\frac{[\sqrt{\lambda}\sqrt{m}-\sqrt{\mu}]^2-\nu m}{\xi}\geq 1+\delta.
\end{equation}
Hence, by applying \eqref{th-aux-ineq1} and \eqref{th-aux-ineq2}, if we assume the condition
$$\sqrt{m}\geq \sqrt{1+\delta}\frac{\sqrt{\mu}+\sqrt{\xi}}{\sqrt{\lambda}-\sqrt{\nu}}
\geq \frac{\sqrt{\mu}+\sqrt{1+\delta}\sqrt{\xi}}{\sqrt{\lambda}-\sqrt{\nu}},$$
then the inequality \eqref{th-success} holds so that we can identify $I_t\subset {\rm supp}X$ by 2-thresholding.

\section*{Appendix G: Proof of Theorem~\ref{lem:num-somp1} and \ref{lem:num-somp2}}

 \renewcommand{\theequation}{G.\arabic{equation}}
\renewcommand{\thetheorem}{G.\arabic{theorem}}
\renewcommand{\thelemma}{G.\arabic{lemma}}
\setcounter{equation}{0} \setcounter{theorem}{0}

In this section, we assume the large system limit such that $\rho$, $\epsilon$, $\alpha$ and $\gamma$ exist. We first need to have the following results.
\begin{theorem}\label{random-dist}\cite{marcenko1967distribution}
Suppose that each entry of $A\in\mathbb{R}^{m\times k}$ is generated from i.i.d. Gaussian random variable $\mathcal{N}(0,1/m)$. Then the probability density of squared singular value of $A$ is given by
\begin{equation}\label{marpas}
d\lambda_{\gamma}(x):=\frac{1}{2\pi\gamma^2}\frac{\sqrt{((1+\gamma)^2-x)(x-(1-\gamma)^2)}}{x}
\end{equation}
where $\gamma=\lim_{n\rightarrow\infty}\sqrt{k/m}$.
\end{theorem}

\begin{corollary}
Suppose that each entry of $A\in\mathbb{R}^{m\times k}$ is generated from i.i.d. Gaussian random variable $\mathcal{N}(0,1/m)$. Then the probability density of singular value of $A$ is given by
\begin{equation}\label{marpas2}
ds_{\gamma}(x):=\frac{1}{\pi\gamma^2}\frac{\sqrt{((1+\gamma)^2-x^2)(x^2-(1-\gamma)^2)}}{x}.
\end{equation}
\end{corollary}
\begin{proof}
This is obtained from Theorem \ref{random-dist} using a simple change of variable.
\end{proof}

\begin{lemma}\label{lem-frob}
Let $r\leq k<m$ be positive integers and $A\in\mathbb{R}^{m\times k}$. Then for any $r$-dimensional subspace $W$ of $R(A)$,  we have
\begin{equation*}
\|A^{*}P_W\|_F^2\geq \sum\limits_{j=1}^r \sigma_{k-j+1}^2(A)
\end{equation*}
where $\sigma_1(A)\geq \sigma_2(A)\geq\cdots\geq \sigma_k(A)\geq 0$.
\end{lemma}
\begin{proof}
Let $A^{*}=\tilde{U}\tilde{\Sigma}\tilde{V}^{*}$ be the extended singular value decomposition of $A^{*}$ where
\begin{eqnarray*}
\tilde{\Sigma}&=&{\rm diag}[\sigma_1,\sigma_2,\cdots,\sigma_m],\\
\tilde{V}&=&[\vb_1,\vb_2,\cdots,\vb_m]
\end{eqnarray*}
and $\sigma_{k+1}=\sigma_{k+2}=\cdots=\sigma_m=0$. If we let $Z=\tilde{V}^{*}P_W$,
then we have
\begin{equation}\label{frob-z}
\|Z\|_F^2={\rm trace}(P_W\tilde{V}\tilde{V}^{*}P_W)={\rm trace}(P_W)=r
\end{equation}
and
\begin{eqnarray*}
\|A^{*}P_W\|_F^2=\|A^{*}\tilde{V}Z\|_F^2=\|\tilde{U}\tilde{\Sigma}Z\|_F^2
\end{eqnarray*}
If we let $Z=[\zb_1^{*},\cdots,\zb_m^{*}]^{*}$, since $W$ is a subspace of $R(A)$ and $R(A)=N(A^{*})^{\perp}$, we have
$$\zb_{k+1}=\zb_{k+2}=\cdots=\zb_m={\bf 0}$$
and
$$\sum\limits_{l=1}^k \|\zb_l\|^2=r.$$
by \eqref{frob-z}.
Since $0\leq\|\zb_l\|^2\leq 1$ for $1\leq j\leq k$, using $\sigma_1(A)\geq \sigma_2(A)\geq \cdots\geq \sigma_k(A)$, we have
\begin{eqnarray*}
\|A^{*}P_W\|_F^2=\|\tilde{U}\tilde{\Sigma}Z\|_F^2=\sum\limits_{l=1}^k \sigma_l^2(A)\|\zb_l\|^2\geq \sum\limits_{j=1}^r\sigma_{k-j+1}^2(A).
\end{eqnarray*}
\end{proof}

\begin{lemma}\label{lem-t}
For $0\leq \gamma<1$ and $0\leq \alpha\leq 1$, we let $0\leq t_{\gamma}(\alpha)\leq 1$ which satisfies $\int_{1-\gamma}^{1-\gamma+2\gamma t_{\gamma}(\alpha)}ds_{\gamma}(x)=\alpha$
where $ds_{\gamma}(x)$ is the probability measure which is given by
$$ds_{\gamma}(x):=\frac{1}{\pi \gamma^2}
\frac{\sqrt{((1+\gamma)^2-x^2)(x^2-(1-\gamma)^2)}}{x}.$$
Then we have for any $0\leq\gamma\leq 1$, $t_{\gamma}(\alpha)\geq t_{1}(\alpha).$
\end{lemma}
\begin{proof}
It is sufficient to show that for any $0\leq t\leq 1$ and $0\leq \gamma<1$,
\begin{eqnarray}\label{lem-ineq1}
\int_{1-\gamma}^{1-\gamma+2\gamma t}ds_{\gamma}(x)\leq
\int_0^{2t} ds_1(x)
\end{eqnarray}
By substituting $s=(x-(1-\gamma))/\gamma$, we have
\begin{eqnarray*}
\int_{1-\gamma}^{1-\gamma+2\gamma t}ds_{\gamma}(x)
=\int_0^{2t}ds_{0,\gamma}(x)
\end{eqnarray*}
where
$$ds_{0,\gamma}(x)=\frac{\sqrt{s}\sqrt{2-s}\sqrt{s+2/\gamma}\sqrt{s+2(1-\gamma)/\gamma}}{\pi(s+(1-\gamma)/\gamma)}.$$
By Lemma \ref{num-intersection}, there is only one root for $ds_{0,\gamma}(x)=ds_1(x)$ in $(0,2)$  and $ds_1(0)>ds_{0,\gamma}(0)$. Then there is some $s_{*}\in (0,2)$ such that
$ds_1(x)>ds_{0,\gamma}(x)$ for $x< s_{*}$ and $ds_1(x)<ds_{0,\gamma}(x)$ for $x> s_{*}$ so that
$$\int_0^{2t}ds_1(x)>\int_0^{2t}ds_{0,\gamma}(x)~{\rm for}~0<2t<s_{*}$$
and
$\int_0^{2t}ds_1(x)-\int_0^{2t}ds_{0,\gamma}(x)$ is a decreasing function on $(s_{*},2)$ such that
$$\int_0^2 ds_1(x)=\int_0^2 ds_{0,\gamma}(x)=1.$$
Hence, for any $t\in (0,1)$,
$$\int_0^{2t}ds_1(x) > \int_0^{2t}ds_{0,\gamma}(x)$$
so that (\ref{lem-ineq1}) holds.
\end{proof}

\begin{lemma}\label{num-intersection}
Let $ds_1(x)$ and $ds_{0,\gamma}(x)$ be probability density functions with support $[0,2]$. Then these probability density functions have only 1 intersection point in (0,2).
\end{lemma}
\begin{proof}
For $s\in (0,2)$
$$\frac{\sqrt{s}\sqrt{2-s}\sqrt{s+2/\gamma}\sqrt{s+2(1-\gamma)/\gamma}}{(s+(1-\gamma)/\gamma)}=\sqrt{4-s^2}$$
if and only if
$$s(s+2/r)(s+2(1-\gamma)/\gamma)=(s+2)(s+(1-\gamma)/\gamma)^2.$$
Expanding both sides, we have
$$(2/\gamma-2)s^2+(4(1-\gamma)/\gamma^2-(1-\gamma)^2/\gamma^2-4(1-\gamma)/\gamma)s-4(1-\gamma)^2/\gamma^2=0$$
so that there is only 1 positive root. If we assume that $ds_{0,\gamma}(x)$ and $ds_1(x)$ have no intersection point, then
$$\int_0^2 ds_1(x)>\int_0^2 ds_{0,\gamma}(x)$$ since $ds_1(0)>ds_{0,\gamma}(0)$. This is a contradiction so that there must be 1 root for $ds_{0,\gamma}(x)=ds_1(x)$ in $(0,2)$.
\end{proof}

{\em Proof of Theorem \ref{lem:num-somp1} and \ref{lem:num-somp2}:}
Note that S-OMP can find $k-r$ correct indices from ${\rm supp}X$ if we have
\begin{equation}\label{erc-somp}
\max\limits_{j\in{\rm supp}X}\|\ab_j^{*}P_{R(P_{R(A_{I_t})}^{\perp}B)}\|^2
>\max\limits_{j\notin{\rm supp}X}\|\ab_j^{*}P_{R(P_{R(A_{I_t})}^{\perp}B)}\|^2
\end{equation}
for each $0\leq t< k-r$, since $\|\ab_j^{*}P_{R(P_{R(A_{I_t})}^{\perp}B)}\|^2=0$ for $j\in {\rm supp}X\cap I_t$. Hence, it is enough to check that the condition \eqref{erc-somp} for $0\leq t<k-r$.

First, for $j\notin {\rm supp}X$, since $\ab_j$ is statistically independent of $P_{R(A_{I_t})}^{\perp}Y$. For $t\leq k-r$, the dimension of $P_{R(A_{I_t})}^{\perp}Y$ is $r$ so that $m\|\ab_j P_{R(P_{R(A_{I_t})}^{\perp}Y)}\|^2$ is of chi-squared distribution of degree of freedom $r$.

On the other hand, for $j\in {\rm supp}X$, we have
\begin{eqnarray*}
\max\limits_{j\in{\rm supp}X}\|\ab_j^{*}P_{R(P_{R(A_{I_t})}^{\perp}B)}\|^2&\geq&
\frac{1}{k}\|A_S^{*}P_{R(P_{R(A_{I_t})}^{\perp}B)}\|_F^2\\
&\geq&\frac{\sum\limits_{j=1}^r \sigma_{k-j+1}^2(A_S)}{k}
\end{eqnarray*}
by using $R(P_{R(A_{I_t})}^{\perp}B)\subset R(A_S)$ and Lemma \ref{lem-frob}, where $A_S$ have singular values
$0<\sigma_k(A_S)\leq \sigma_{k-1}(A_S)\leq\cdots\leq \sigma_1(A_S)$. If we let
$$ds_{\gamma}(x):=\frac{1}{\pi\gamma^2}\frac{\sqrt{((1+\gamma)^2-x^2)(x^2-(1-\gamma)^2)}}{x}$$
then by (\ref{marpas}), we have
\begin{eqnarray}\label{mpd1}
\lim\limits_{n\rightarrow\infty}\frac{\sum\limits_{j=1}^r \sigma_{k-j+1}^2(A_S)}{k}=
\int_{(1-\gamma)^2}^{(1-\gamma+2\gamma t_\gamma(\alpha))^2}xd\lambda_{\gamma}(x)
\end{eqnarray}
where $0\leq t_{\gamma}(\alpha)\leq 1$ is the value satisfying
$$\int_{1-\gamma}^{1-\gamma+2\gamma t_{\gamma}(\alpha)}ds_{\gamma}(x)=\alpha=\lim\limits_{n\rightarrow\infty}\frac{r}{k}$$
Using Lemma \ref{lem-t}, we have $t_{\gamma}(\alpha)\geq t_1(\alpha)$ for all $0\leq \alpha\leq 1$ and $0<\gamma\leq 1$. Using this and \eqref{mpd1}, we have
\begin{eqnarray}\label{mpd2}
\notag&&\int_{(1-\gamma)^2}^{(1-\gamma+2\gamma t_\gamma(\alpha))^2}xd\lambda_{\gamma}(x)\\
\notag&\geq&\int_{(1-\gamma)^2}^{(1-\gamma+2\gamma t_1(\alpha))^2}xd\lambda_{\gamma}(x)\\
\notag&\geq&\int_0^{4 t_1(\alpha)^2}
[(1-\gamma)^2+\gamma s]d\lambda_1(s)\\
&=&(1-\gamma)^2\alpha+\gamma\int_0^{4t_1(\alpha)^2}sd\lambda_1(s)
\end{eqnarray}
where we used the integral by substitution with $s=(x-(1-\gamma)^2)/\gamma$ and the inequality $(1-\gamma+2\gamma t)^2-(1-\gamma)^2\geq 4\gamma t^2$ for $0\leq t\leq 1$. Substituting (\ref{mpd2}) into (\ref {mpd1}), we have
\begin{eqnarray}\label{j-supp-lower}
\notag \liminf\limits_{n\rightarrow\infty}\max\limits_{j\in{\rm supp}X}\|\ab_j^{*}P_{R(P_{R(A_{I_t})}^{\perp}B)}\|^2&\geq&\liminf\limits_{n\rightarrow\infty}\frac{\sum\limits_{j=1}^r \sigma_{k-j+1}^2(A_S)}{k}\\
\notag &\geq&\alpha\left[(1-\gamma)^2+\gamma \frac{\int_0^{4t_1(\alpha)^2}sd\lambda_1(s)}{\alpha}\right]\\
&=&\lim\limits_{n\rightarrow\infty}\frac{r}{m}(1/\gamma-1)^2+\alpha\gamma F(\alpha)
\end{eqnarray}
where $F(\alpha):=(1/\alpha)\int_0^{4t_1(\alpha)^2}sd\lambda_1(s)$ is an increasing function with respect to $\alpha$ such that $\lim_{\alpha\rightarrow 0}F(\alpha)=0$ and $\alpha(1)=1$, and $\alpha\gamma^2=(\lim_{n\rightarrow\infty}r/k)(\lim_{n\rightarrow\infty}k/m)=\lim_{n\rightarrow\infty}r/m$.

For noisy measurement $Y$, we have the following inequality:
\begin{eqnarray}\label{perturb-diff}
\notag & &\left|\|P_{R(P_{R(A_{I_t})}^{\perp}Y)}\ab_j\|^2-\|P_{R(P_{R(A_{I_t})}^{\perp}B)}\ab_j\|^2\right|\\
\notag &\leq&(\|P_{R(P_{R(A_{I_t})}^{\perp}Y)}\ab_j\|+\|P_{R(P_{R(A_{I_t})}^{\perp}B)}\ab_j\|)\left|\|P_{R(P_{R(A_{I_t})}^{\perp}Y)}\ab_j\|-\|P_{R(P_{R(A_{I_t})}^{\perp}B)}\ab_j\|\right|\\
\notag &\leq&2\|\ab_j\|\|P_{R(P_{R(A_{I_t})}^{\perp}Y)}\ab_j-P_{R(P_{R(A_{I_t})}^{\perp}B)}\ab_j\|\\
\notag &\leq&2\|\ab_j\|\|P_{R(Y)}\ab_j-P_{R(B)}\ab_j\|\leq 2\|\ab_j\|^2\|P_{R(Y)}-P_{R(B)}\|\\&&\longrightarrow 2\|P_{R(Y)}-P_{R(B)}\|\leq
\frac{4(\sigma_{\max}(B)+\sigma_{\min}(B))\|N\|}{\sigma_{\min}(B)(\sigma_{\min}(B)-\|N\|)}=\frac{4(\kappa(B)+1)}{{\sf SNR}_{\min}(B)-1}
\end{eqnarray}
as $n\rightarrow\infty$, where ${\sf SNR}_{\min}(B)=\sigma_{\min}(B)/\|N\|$.

\indent Then we consider two limiting cases according to the number of measurement vectors.\\
{\bf  (Case 1 : Theorem \ref{lem:num-somp1})} For $t\leq k-r$, $\{m\|\ab_j^{*}P_{R(P_{R(A_{I_t})}^{\perp}B)}\|^2:j\notin {\rm supp}X\}$ are independent chi-squared random variables of degree of freedom $r$ so that by Lemma \ref{chi-max}, we have
\begin{equation}\label{gamma1}
\lim_{n\rightarrow\infty}\max\limits_{j\notin {\rm supp}X}\frac{m\|\ab_j^{*}P_
{R(P_{R(A_{I_t})}^{\perp}B)}\|^2}{2\log{(n-k)}}=1.
\end{equation}
Here we assume that
\begin{equation}\label{snr-somp-rfixed}
{\sf SNR}_{\min}(Y)>1+4\frac{k}{r}(\kappa(B)+1)
\end{equation}
and
\begin{equation}\label{num-somp-rfixed}
m>k\left[1-\frac{4k}{r}\frac{\kappa(B)+1}{{\sf SNR}_{\min}-1}\right]^{-1}
2(1+\delta)\frac{\log{(n-k)}}{r}.
\end{equation}
Then by Mar\'{c}enko-Pastur theorem \cite{marcenko1967distribution},
$$\lim\limits_{n\rightarrow\infty}\sigma_{\min}(A_S)=\lim\limits_{n\rightarrow\infty}
(1-\sqrt{k/m})^2\geq \lim\limits_{n\rightarrow\infty}\left(1-\sqrt{r/(2\log{(n-k)})}\right)^2=1$$
so that
\begin{eqnarray}\label{j-supp-lower2}
\notag \liminf\limits_{n\rightarrow\infty}\max\limits_{j\in{\rm supp}X}\frac{m\|\ab_j^{*}P_{R(P_{R(A_{I_t})}^{\perp}B)}\|^2}{2\log{(n-k)}}&\geq&\liminf\limits_{n\rightarrow\infty}\frac{m}{2\log{(n-k)}}\frac{\sum\limits_{j=1}^r \sigma_{k-j+1}^2(A_S)}{k}\\&\geq& \liminf\limits_{n\rightarrow\infty}\frac{r}{2\log{(n-k)}}
\left(\frac{1}{\gamma}\right)^2.
\end{eqnarray}
Combining \eqref{perturb-diff} and \eqref{j-supp-lower2}, for noisy measurement $Y$, we have
\begin{eqnarray*}
&&\liminf\limits_{n\rightarrow\infty}\max\limits_{j\in {\rm supp}X}
\frac{m\|\ab_j^{*}P_{R(P_{R(A_{I_t})}^{\perp}Y)}\|^2}{2\log{(n-k)}}\\
&\geq&\liminf\limits_{n\rightarrow\infty}
\frac{m\left[\|\ab_j^{*}P_{R(P_{R(A_{I_t})}^{\perp}B)}\|^2-\left|
\|\ab_j^{*}P_{R(P_{R(A_{I_t})}^{\perp}Y)}\|^2-|\ab_j^{*}P_{R(P_{R(A_{I_t})}^{\perp}B)}\|^2\right|\right]}{2\log{(n-k)}}\\
&\geq&\liminf\limits_{n\rightarrow\infty}\frac{r}{2\log{(n-k)}}\left[1-\frac{4k}{r}
\frac{\kappa(B)+1}{{\sf SNR}_{\min}(Y)-1}\right]\frac{1}{\gamma^2}\geq 1+\delta
\end{eqnarray*}
if we have \eqref{num-somp-rfixed}. Hence, when $r$ is a fixed number, if we have \eqref{num-somp-rfixed}, then we can identify $k-r$ correct indices of ${\rm supp}X$ with subspace S-OMP in LSMMV.
\\
\indent {\bf (Case 2: Theorem \ref{lem:num-somp2})} Similarly as in the previous case, for $t< k-r$, $\{m\|\ab_j^{*}P_
{R(P_{R(A_{I_t})}^{\perp}B)}\|^2:j\notin {\rm supp}X\}$ are independent chi-squared  distribution.  Since $\lim_{n\rightarrow\infty}(\log{n})/r=0$, by Lemma 3 in \cite{fletcher2009necessary}, we have
\begin{equation}\label{gamma2}
\lim_{n\rightarrow\infty}\max\limits_{j\notin {\rm supp}X}\frac{m\|\ab_j^{*}P_
{R(P_{R(A_{I_t})}^{\perp}B)}\|^2}{r}=1.
\end{equation}
By using \eqref{perturb-diff}, we have
\begin{eqnarray}\label{omp-supp2}
\notag &&\liminf_{n\rightarrow\infty}\max\limits_{j\in {\rm supp}X}\frac{m\|\ab_j^{*}P_
{R(P_{R(A_{I_t})}^{\perp}Y)}\|^2}{r}\\
&\geq&\left(\frac{1}{\gamma}-1\right)^2+F(\alpha)
\frac{1}{\gamma}-\frac{4}{\alpha}\frac{\kappa(B)+1}{{\sf SNR}_{\min}(B)-1}\frac{1}{\gamma^2}.
\end{eqnarray}
We let
\begin{equation}\label{snr-somp2}
{\sf SNR}_{\min}(B)>1+\frac{4}{\alpha}(\kappa(B)+1)
\end{equation}
and
\begin{equation}\label{num-somp2}
m>k(1+\delta)^2\frac{1}{\left(1-\frac{4}{\alpha}\frac{(\kappa(B)+1)}{{\sf SNR}_{\min}-1}\right)^2}\left[2-F(\alpha)\right]^2
\end{equation}
for some $\delta>0$. Note that \eqref{num-somp2} is equivalent to
\begin{equation*}
\frac{1}{\gamma}>(1+\delta)\frac{1}{1-\frac{4}{\alpha}\frac{(\kappa(B)+1)}{{\sf SNR}_{\min}-1}}[2-F(\alpha)]
\end{equation*}
Again we let
$$u:=F(\alpha)~{\rm and}~v:=\frac{4}{\alpha}\frac{\kappa(B)+1}{{\sf SNR}_{\min}(B)-1}.$$
Then for a quadratic function $Q(x)=(x-1)^2+ux-vx^2$, if $x>(1+\delta)(2-u)/(1-v)$, then we have
\begin{eqnarray}\label{aux-ineq2}
Q(x)&=&(1-v)x^2-(2-u)x+1=(1-v)x\left[x-\frac{2-u}{(1-v)}\right]+1\\
&>&\delta(1+\delta)\frac{(2-u)^2}{1-v}+1\geq  1+\delta(1+\delta)
\end{eqnarray}
since $1-v>0$ by \eqref{snr-somp2} and $0\leq u\leq 1$. Combining \eqref{omp-supp2} and \eqref{aux-ineq2}, we have for $0\leq t<k-r$ and $j\in {\rm supp}X$, we have
$$\liminf_{n\rightarrow\infty}\max\limits_{j\in {\rm supp}X}\frac{m\|\ab_j^{*}P_
{R(P_{R(A_{I_t})}^{\perp}Y)}\|^2}{r}\geq 1+\delta(1+\delta)$$
for some $\delta>0$. Hence, in the case of $\lim_{n\rightarrow\infty} r/k=\alpha>0$, we can identify the correct indices of ${\rm supp}X$ if we have (\ref{num-somp2}).

\section*{Acknowledgments}
 This research was  supported by the Korea Science and Engineering Foundation
(KOSEF) grant funded by the Korean government (MEST) (No.
2009-0081089).  The authors would like to thank Dr. Dmitry Malioutov for providing the $l_{1,2}$ mixed norm code in \cite{malioutov2005ssr}.

\bibliographystyle{IEEEbib}
\bibliography{bispl2010}

\newpage

\end{document}